\documentclass[11pt]{article}%
\usepackage{amsmath}
\usepackage{graphicx}
\usepackage{amsfonts}
\usepackage{amssymb}%
\newcommand{\eqnb}{\begin{equation}}
\newcommand{\eqne}{\end{equation}}

\newtheorem{The}{Theorem}

\newtheorem{Lem}{Lemma}
\newtheorem{Pro}{Proposition}
\newtheorem{Rem}{Remark}
\begin{document}

\title{The Chaos of Propagation in a Retrial Supermarket Model}
\author{Quan-Lin Li${}^{1}$ \hspace{0.1in}Meng Wang${}^{1}$\hspace{0.1in}John C.S.
Lui${}^{2}$ \hspace{0.1in}Yang Wang ${}^{3}$\\${}^{1}$ School of Economics and Management Sciences \\Yanshan University, Qinhuangdao 066004, China \\${}^{2}$ Department of Computer Science \& Engineering \\The Chinese University of Hong Kong, Shatin, N.T, Hong Kong \\${}^{3}$ Institute of Network Computing \& Information Systems\\Peking University, China }
\maketitle

\begin{abstract}
When decomposing the total orbit into $N$ sub-orbits (or simply orbits)
related to each of $N$ servers and through comparing the numbers of customers
in these orbits, we introduce a retrial supermarket model of $N$ identical
servers, where two probing-server choice numbers are respectively designed for
dynamically allocating each primary arrival and each retrial arrival into
these orbits when the chosen servers are all busy. Note that the designed
purpose of the two choice numbers can effectively improve performance measures
of this retrial supermarket model.

This paper analyzes a simple and basic retrial supermarket model of $N$
identical servers, that is, Poisson arrivals, exponential service and retrial
times. To this end, we first provide a detailed probability computation to set
up an infinite-dimensional system of differential equations (or mean-field
equations) satisfied by the expected fraction vector. Then, as $N\rightarrow
\infty$, we apply the operator semigroup to obtaining the mean-field limit (or
chaos of propagation) for the sequence of Markov processes which express the
state of this retrial supermarket model. Specifically, some simple and basic
conditions for the mean-field limit as well as for the Lipschitz condition are
established through the first two moments of the queue length in any orbit.
Finally, we show that the fixed point satisfies a system of nonlinear
equations which is an interesting networking generalization of the tail
equations given in the M/M/1 retrial queue, and also use the fixed point to
give performance analysis of this retrial supermarket model through numerical
computation. Noting that there are few available works on the analysis of
retrial queueing networks in the current literature, we believe the mean-field
method given in this paper can open a new avenue in the future study of
retrial supermarket models, and more generally, of retrial queueing networks.

\vskip                              0.5cm


\end{abstract}

\section{Introduction}

Retrial queues are an important mathematical model for studying telephone
switch systems, digital cellular mobile networks, computer networks and so on.
During the last two decades, considerable attention has been paid to the study
of retrial queues. Readers may refer to, for example, six survey papers by
Yang and Templeton \cite{Yang:1987}, Falin \cite{Fal:1990}, Kulkarni and Liang
\cite{Kul:1997}, Artalejo \cite{Art:1999, Art:2010} and G\'{o}mez-Corral
\cite{Gom:2006}, and three books by Falin and Templeton \cite{Fal:1997},
Artalejo and G\'{o}mez-Corral \cite{Art:2008} and Li \cite{Li:2010}.

Few available works have been done on the analysis of retrial queueing
networks in the current literature. Kim \cite{Kim:2010} applied the fluid
limit to considering the stability of a retrial queueing network with
different classes of customers and restricted resource pooling. Avrachenkov et
al \cite{Avr:2013} discussed a retrial queue with two input streams and two
orbits, and derived the stationary joint distribution of the numbers of
customers in the two orbits by means of the two-dimensional probability
generating functions. This paper provides a mean-field method to study a
large-scale system of $N$ parallel retrial queues under a dynamic randomized
load balancing scheme. Also, this paper examines the performance of this
large-scale system by means of some numerical computation.

Dynamic randomized load balancing is often referred to as the supermarket
model. Recently, some supermarket models have been analyzed by means of
queueing theory as well as Markov processes. For the simple supermarket model
with Poisson inputs and exponential service times, Vvedenskaya et al
\cite{Vve:1996} applied the operator semigroups to providing a mean-field
limit for the sequence of Markov processes. Mitzenmacher \cite{Mit:1996} also
analyzed the same supermarket model in terms of the density-dependent jump
Markov processes. Turner \cite{Tur:1996, Tur:1998} provided a martingale
approach which can simplify some crucial discussion of this supermarket model.
Furthermore, Graham \cite{Gra:2000, Gra:2004} studied the path space evolution
and showed that starting from independent initial states, as $N\rightarrow
\infty$ the queues of the limiting process evolve independently. Luczak and
McDiarmid \cite{Luc:2006, Luc:2007} showed that the length of the longest
queue scales as $(\log\log N)/\log d+O(1)$. From the modeling point of view,
certain generalizations of the simple supermarket model have been explored by
studying several variations, important examples include Vvedenskaya and Suhov
\cite{Vve:1997}, Mitzenmacher \cite{Mit:1999}, Foss and Chernova
\cite{Foss:1998}, Bramson \cite{Bra:2011}, Bramson et al \cite{Bra:2010,
Bra:2012, Bra:2013}, MacPhee et al \cite{Mac:2012}, Li \cite{Li:2011}, Li et
al \cite{Li:2011a, Li:2013}, Martin and Suhov \cite{Mar:1999}, Martin
\cite{Mar:2001} and Suhov and Vvedenskaya \cite{Suh:2002}.

The mean-field theory always plays an important role in the study of
supermarket models. Readers may refer to recent publications for the
mean-field models, among which are Dawson \cite{Daw:1983}, Sznitman
\cite{Szn:1989}, Vvedenskaya and Suhov \cite{Vve:1997}, Dawson et al
\cite{Daw:2005}, Le Boudec et al \cite{Bou:2007}, Bordenave et al
\cite{Bor:2009}, Gast and Gaujal \cite{Gast:2009, Gast:2012}, Gast et al
\cite{Gast:2011} and Tsitsiklis and Xu \cite{Tsi:2012}. Readers may also refer
to Mitzenmacher \cite{Mit:1999} and Benaim and Le Boudec \cite{Ben:2008} for
two excellent surveys of many interesting mean-field models used in practice.

The main contributions of this paper are threefold. The first one is to
introduce a simple and basic retrial supermarket model of $N$ identical
servers with Poisson arrivals, exponential service and retrial times and with
two probing-server choice numbers. To analyze this retrial supermarket model,
we provide a detailed probability computation to set up a system of
differential equations (or mean-field equations) satisfied by the expected
fraction vector through observing five crucial modeling cases. The second
contribution is to use the operator semigroup to provide the mean-field limit
(or chaos of propagation) for the sequence of Markov processes, where the
Lipschitz condition is also established in order to prove the existence and
uniqueness of the solution to the infinite-dimensional system of limiting
differential equations through the Picard approximation. Crucially, some
simple and basic conditions for the mean-field limit as well as for the
Lipschitz condition are well organized through the first two moments of the
queue length in any orbit. The third contribution is to compute the fixed
point by means of a system of nonlinear equations which is an interesting
networking generalization of the tail equations given in the M/M/1 retrial
queue, and also use the fixed point to give performance analysis of this
retrial supermarket model through some numerical computation.

The remainder of this paper is organized as follows. In Section 2, we describe
a simple and basic retrial supermarket model of $N$ identical servers with
Poisson arrivals, exponential service and retrial times and with two different
probing-server choice numbers. We then use the fraction vector to construct an
infinite-dimensional Markov process which expresses the state of the retrial
supermarket model. In Section 3, we provide a detailed probability computation
to set up a system of differential equations satisfied by the expected
fraction vector. In Section 4, we establish the Lipschitz condition whose aim
is to prove the existence and uniqueness of the solution to the
infinite-dimensional system of limiting differential equations by means of the
Picard approximation. In Section 5, we apply the operator semigroup to
providing a mean-field limit for the sequence of Markov processes. In Section
6, we compute the fixed point by means of a system of nonlinear equations, and
also give performance analysis of this retrial supermarket model through some
numerical computation. Some concluding remarks are given in Section 7.

\section{A Retrial Supermarket Model}

In this section, we first describe a simple and basic retrial supermarket
model of $N$ identical servers with Poisson arrivals, exponential service and
retrial times and with two probing-server choice numbers. Then we use the
fraction vector to construct an infinite-dimensional Markov process which
expresses the state of the retrial supermarket model.

\subsection{Model description}

We describe the retrial supermarket model of $N$ identical servers as follows.
Primary customers arrive at the queueing system as a Poisson process with
arrival rate $N\lambda$, and the service times at each server are i.i.d.
exponentially random variables with service rate $\mu$. There is no waiting
space but there is an orbit of infinite size corresponding to each server.
Each primary arriving customer chooses $d_{1}\geq1$ servers independently and
uniformly at random from the $N$ servers. If there exist some idle servers in
the $d_{1}$ chosen servers, then it enters one idle server with the fewest
customers in the orbit and receives service immediately; otherwise it enters
the orbit of one (busy) server with the fewest customers in the orbit, and
makes a retrial at a later time. Similarly, each retrial arriving customer
chooses $d_{2}\geq1$ servers independently and uniformly at random from the
$N$ servers, if there exist some idle servers in the $d_{2}$ chosen servers,
then it enters one idle server with the fewest customers in the orbit and
receives service immediately; otherwise it comes back to its original orbit
again. Retrial customers behave independently of each other and are persistent
in the sense that they keep making retrials until they receive their requested
service. Successive inter-retrial times of each customer in these orbits are
exponential with retrial rate $\theta$. We assume that all the random
variables defined above are independent of each other, and that this queueing
system is operating in the region $\rho=\lambda/\mu<1$. Figure 1 provides a
physical interpretation for this retrial supermarket model.

\begin{figure}[ptbh]
\centering
\includegraphics[width=12cm]{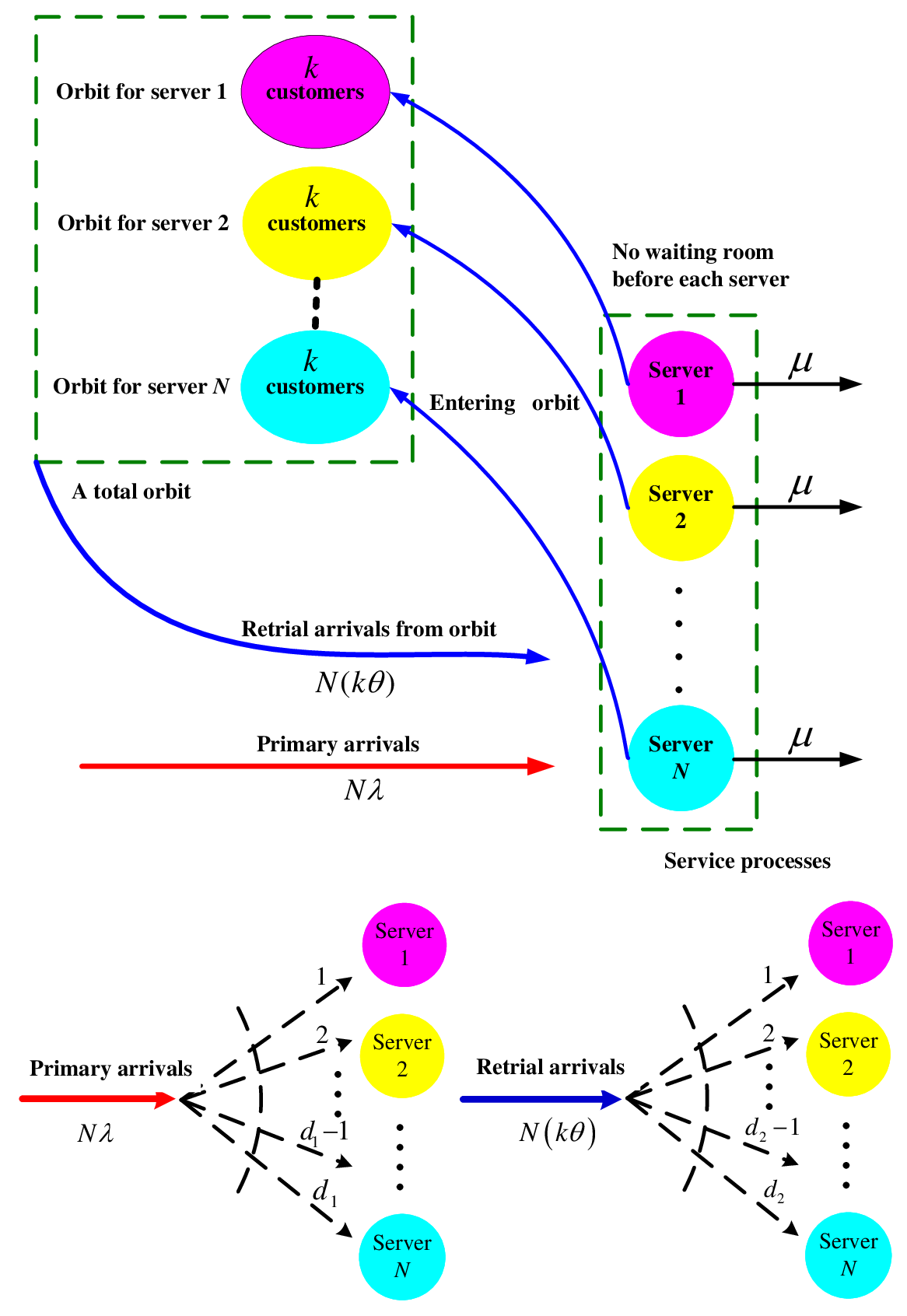} \caption{A physical illustration for
the retrial supermarket model}%
\label{figure: fig-1}%
\end{figure}

From Figure 1, it is seen that the total orbit is decomposed into the $N$
orbits corresponding to each of the $N$ servers, while such a decomposition
will not change the retrial behavior of customers in the total orbit because
of the following equation of retrial rates:%
\[
\theta\sum_{j=1}^{N}k_{j}=\sum_{j=1}^{N}k_{j}\theta,
\]
where $\theta\sum_{j=1}^{N}k_{j}$ is the total retrial rate from the total
orbit; while $k_{j}\theta$ is the retrial rate from the $j$th orbit when
$k_{j}$ is the number of customers in the $j$th orbit.

Based on this decomposition of the total orbit into the $N$ orbits, we can
introduce the two probing-server choice numbers $d_{1}$ and $d_{2}$ for
dynamically allocating a primary or retrial arriving customer into its
suitable orbit. Note that in this retrial supermarket model, the $N$ retrial
queues are symmetric and exchangeable, and they also depend upon each other
through the dynamical randomized load balancing scheme, thus we can study the
mean-field limit and show that the two choice numbers can effectively improve
the performance of this system including the stationary queue length mean, the
expected sojourn time, and the stationary throughput. On the other hand, the
dynamical randomized load balancing scheme $\left(  d_{1},d_{2}\right)  $ can
be realized very well through the present Internet and information
technologies including data centers, big data and RFID.

The following lemma provides a sufficient condition under which this retrial
supermarket model of $N$ identical servers is stable. This proof can be given
by means of a similar coupling method to Theorems 4 and 5 in Martin and Suhov
\cite{Mar:1999}.

\begin{Lem}
For the retrial supermarket model of $N$ identical servers with Poisson
arrivals, exponential service and retrial times and with two different
probing-server choice numbers $d_{1}\geq1$ and $d_{2}\geq1$, it is stable if
$\rho=\lambda/\mu<1$.
\end{Lem}

\textbf{Proof:} \ If $d_{1}=d_{2}=1$, then this retrial supermarket model of
$N$ identical servers is equivalent to a system of $N$ independent M/M/1
retrial queues with exponential retrial times. From Chapter 1 of Artalejo and
G\'{o}mez-Corral \cite{Art:2008}, it is seen that the M/M/1 retrial queues
with exponential retrial times is stable if $\rho<1$. Using a coupling method,
as given in Theorems 4 and 5 of Martin and Suhov \cite{Mar:1999}, it is clear
that for a fixed number $N=1,2,3,\ldots$, this retrial supermarket model of
$N$ identical servers is stable if $\rho<1$. This completes the proof.
\textbf{{\rule{0.08in}{0.08in}}}

\subsection{An infinite-dimensional Markov Process}

For $k\geq0$, we denote by $n_{k}^{\left(  W\right)  }\left(  t\right)  $ and
$n_{k}^{\left(  I\right)  }\left(  t\right)  $ the numbers of busy servers and
of idle servers\ with at least $k$ customers in the orbit at time $t$.
Clearly, $n_{0}^{\left(  W\right)  }\left(  t\right)  +n_{0}^{\left(
I\right)  }\left(  t\right)  =N$ and $0\leq n_{k}^{\left(  W\right)  }\left(
t\right)  ,n_{k}^{\left(  I\right)  }\left(  t\right)  \leq N$ for $k\geq0$.

For $k\geq0$, we write%
\[
U_{W,k}^{\left(  N\right)  }\left(  t\right)  =\frac{n_{k}^{\left(  W\right)
}\left(  t\right)  }{N}%
\]
and%
\[
U_{I,k}^{\left(  N\right)  }\left(  t\right)  =\frac{n_{k}^{\left(  I\right)
}\left(  t\right)  }{N},
\]
which are the fractions of busy servers and of idle servers with at least $k$
customers in the orbit\ at time $t$, respectively. Let
\[
U_{k}^{\left(  N\right)  }\left(  t\right)  =\left(  U_{W,k}^{\left(
N\right)  }\left(  t\right)  ,U_{I,k}^{\left(  N\right)  }\left(  t\right)
\right)
\]
and%
\[
U^{\left(  N\right)  }\left(  t\right)  =\left(  U_{0}^{\left(  N\right)
}\left(  t\right)  ,U_{1}^{\left(  N\right)  }\left(  t\right)  ,U_{2}%
^{\left(  N\right)  }\left(  t\right)  ,\ldots\right)  .
\]
Then the state of the retrial supermarket model can be described as the random
vector $U^{\left(  N\right)  }\left(  t\right)  $ for $t\geq0$. Since the
arrival process to this queueing system is Poisson and the service and retrial
times of each customer are all exponential, $\left\{  U^{\left(  N\right)
}\left(  t\right)  ,t\geq0\right\}  $ is an infinite-dimensional Markov
process whose state space is given by%
\begin{align*}
\Omega_{N}=  &  \left\{  \left(  u_{0}^{\left(  N\right)  },u_{1}^{\left(
N\right)  },u_{2}^{\left(  N\right)  },\ldots\right)  :\left(  1,1\right)
\geq u_{k}^{\left(  N\right)  }\geq u_{k+1}^{\left(  N\right)  }\geq
0\text{,}\text{ and }Nu_{k}^{\left(  N\right)  }\text{ is a}\right. \\
&  \text{ }\left.  \text{two-dimensional row vector of nonnegative integers
for }k\geq0\right\}  .
\end{align*}

For a fixed pair array $\left(  t,N\right)  $ with $t\geq0$ and
$N=1,2,3,\ldots$, it is easy to see from the stochastic order that
$U_{I,k}^{\left(  N\right)  }\left(  t\right)  \geq U_{I,k+1}^{\left(
N\right)  }\left(  t\right)  $ and $U_{W,k}^{\left(  N\right)  }\left(
t\right)  \geq U_{W,k+1}^{\left(  N\right)  }\left(  t\right)  $\ for $k\geq
0$. This gives
\begin{equation}
\left(  1,1\right)  \geq U_{0}^{\left(  N\right)  }\left(  t\right)  \geq
U_{1}^{\left(  N\right)  }\left(  t\right)  \geq U_{2}^{\left(  N\right)
}\left(  t\right)  \geq\cdots\geq0. \label{Equat0}%
\end{equation}

To study the infinite-dimensional Markov process $\left\{  U^{\left(
N\right)  }\left(  t\right)  :t\geq0\right\}  $, we need to consider an
expected fraction vector. For $k\geq0$, we write%
\[
u_{W,k}^{\left(  N\right)  }\left(  t\right)  =E\left[  U_{W,k}^{\left(
N\right)  }\left(  t\right)  \right]
\]
and%
\[
u_{I,k}^{\left(  N\right)  }\left(  t\right)  =E\left[  U_{I,k}^{\left(
N\right)  }\left(  t\right)  \right]  .
\]
Let%
\[
u_{k}^{\left(  N\right)  }\left(  t\right)  =\left(  u_{W,k}^{\left(
N\right)  }\left(  t\right)  ,u_{I,k}^{\left(  N\right)  }\left(  t\right)
\right)
\]
and%
\[
\mathbf{u}^{\left(  N\right)  }\left(  t\right)  =\left(  u_{0}^{\left(
N\right)  }\left(  t\right)  ,u_{1}^{\left(  N\right)  }\left(  t\right)
,u_{2}^{\left(  N\right)  }\left(  t\right)  ,\ldots\right)  .
\]

\section{The System of Differential Equations}

In this section, we provide a detailed probability computation to set up a
system of differential equations satisfied by the expected fraction vector
$\mathbf{u}^{\left(  N\right)  }\left(  t\right)  $. Note that the probability
computation is given through observing five crucial modeling cases with
respect to the arrival, service and retrial processes in this retrial
supermarket model.

\subsection{The system of differential equations}

\begin{figure}[ptbh]
\centering      \includegraphics[width=10cm]{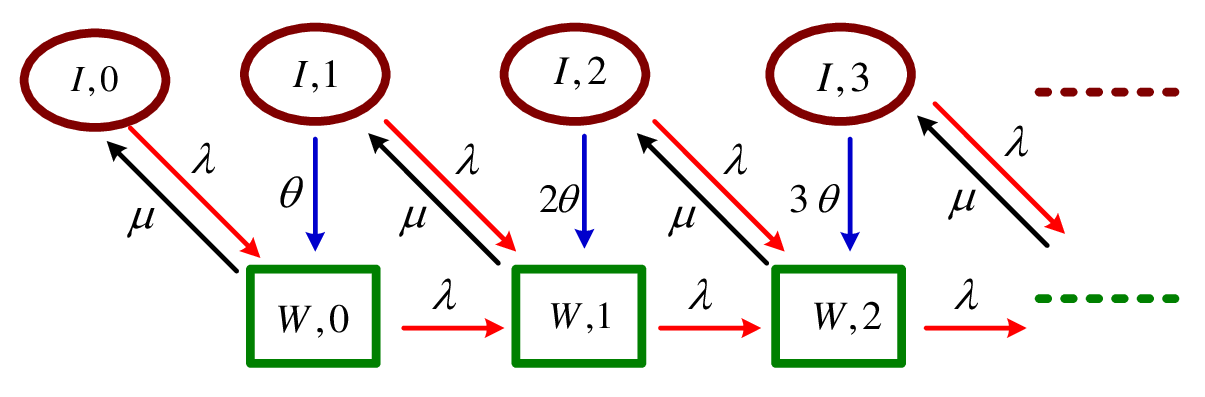} \caption{The state
transitions of any retrial queue in this retrial supermarket model}%
\label{figure: fig-2}%
\end{figure}

Noting that Figure 2 is the state transitions of any retrial queue, the $N$
retrial queues in this retrial supermarket model must be $N$ copies of the
state transition relation given in Figure 2. Based on this, we first set up a
differential equation satisfied by the expected fraction $u_{W,k}^{\left(
N\right)  }\left(  t\right)  $ through observing\ five types of changes with
respect to the expected fraction of the busy servers over a small time period
$[0,$d$t)$. To this end, we need to consider the five different cases as follows.

\textbf{Case one (Primary arrivals and busy servers):} If the $d_{1}$\ chosen
servers are all busy and each orbit of the $d_{1}$\ chosen servers contains at
least $k-1$ customers, then any primary arriving customer must join one orbit
with $k-1$ customers. In this case, the rate that if the $d_{1}$\ chosen
servers are all busy and each of their orbits has at least $k-1$ customers,
any primary arriving customer must join one orbit with $k-1$ customers is
given by%
\begin{align}
&  N\lambda\left[  u_{W,k-1}^{\left(  N\right)  }\left(  t\right)
-u_{W,k}^{\left(  N\right)  }\left(  t\right)  \right]  \sum_{m=1}^{d_{1}%
}C_{d_{1}}^{m}\left[  u_{W,k-1}^{\left(  N\right)  }\left(  t\right)
-u_{W,k}^{\left(  N\right)  }\left(  t\right)  \right]  ^{m-1}\left[
u_{W,k}^{\left(  N\right)  }\left(  t\right)  \right]  ^{d_{1}-m}%
\text{d}t\nonumber\\
&  =N\lambda\left\{  \sum_{m=0}^{d_{1}}C_{d_{1}}^{m}\left[  u_{W,k-1}^{\left(
N\right)  }\left(  t\right)  -u_{W,k}^{\left(  N\right)  }\left(  t\right)
\right]  ^{m}\left[  u_{W,k}^{\left(  N\right)  }\left(  t\right)  \right]
^{d_{1}-m}-\left[  u_{W,k}^{\left(  N\right)  }\left(  t\right)  \right]
^{d_{1}}\right\}  \text{d}t\nonumber\\
&  =N\lambda\left[  \left(  u_{W,k-1}^{\left(  N\right)  }\left(  t\right)
\right)  ^{d_{1}}-\left(  u_{W,k}^{\left(  N\right)  }\left(  t\right)
\right)  ^{d_{1}}\right]  \text{d}t, \label{Equ-1}%
\end{align}
where $C_{d_{1}}^{m}=d_{1}!/\left[  m!\left(  d_{1}-m\right)  \right]  $ is a
binomial coefficient, $\left[  u_{W,k-1}^{\left(  N\right)  }\left(  t\right)
-u_{W,k}^{\left(  N\right)  }\left(  t\right)  \right]  ^{m}$ is the
probability that any primary arriving customer who can only choose and enter
one orbit makes $m$ independent selections among the $m$ selected orbits with
$k-1$ customers at time $t$, and $\left[  u_{W,k}^{\left(  N\right)  }\left(
t\right)  \right]  ^{d_{1}-m}$ is the probability that the $d_{1}-m$ chosen
servers are busy, and each of their orbits has at least $k$ customers at time
$t$.

\textbf{Case two (Primary arrivals and idle servers):} If there exist some
idle servers among the $d_{1}$\ chosen servers and each orbit of the $d_{1}%
$\ chosen servers contains at least $k$ customers, then any primary arriving
customer enters one idle server whose orbit contains the fewest $j$ customers
among all the idle servers for $j\geq k$, and then receives service
immediately. Clearly, each orbit of the idle servers contains at least $j$
customers. To observe the $d_{1}$\ chosen servers, Figure 3 provides a
classification of the $d_{1}$\ chosen servers by means of the states of
servers as well as the numbers of customers in the orbits. In this case, the
corresponding rate is given by%
\begin{align*}
&  N\lambda\sum_{j=k}^{\infty}\left[  u_{I,j}^{\left(  N\right)  }\left(
t\right)  -u_{I,j+1}^{\left(  N\right)  }\left(  t\right)  \right]
\sum\limits_{\substack{m_{1},m_{2},m_{3}\geq0\\m_{1}+m_{2}+m_{3}=d_{1}%
-1}}\left(
\begin{array}
[c]{c}%
d_{1}-1\\
m_{1},m_{2},m_{3}%
\end{array}
\right)  \left[  u_{I,j}^{\left(  N\right)  }\left(  t\right)  -u_{I,j+1}%
^{\left(  N\right)  }\left(  t\right)  \right]  ^{m_{1}}\\
&  \times\left[  u_{I,j+1}^{\left(  N\right)  }\left(  t\right)  \right]
^{m_{2}}\left[  u_{W,k}^{\left(  N\right)  }\left(  t\right)  \right]
^{m_{3}}\text{d}t=N\lambda\sum_{j=k}^{\infty}\left[  u_{I,j}^{\left(
N\right)  }\left(  t\right)  -u_{I,j+1}^{\left(  N\right)  }\left(  t\right)
\right]  \left[  u_{I,j}^{\left(  N\right)  }\left(  t\right)  +u_{W,k}%
^{\left(  N\right)  }\left(  t\right)  \right]  ^{d_{1}-1}\text{d}t,
\end{align*}
where%
\[
\left(
\begin{array}
[c]{c}%
d_{1}-1\\
m_{1},m_{2},m_{3}%
\end{array}
\right)  =\frac{\left(  d_{1}-1\right)  !}{m_{1}!m_{2}!m_{3}!}.
\]

\begin{figure}[ptbh]
\centering      \includegraphics[width=9cm]{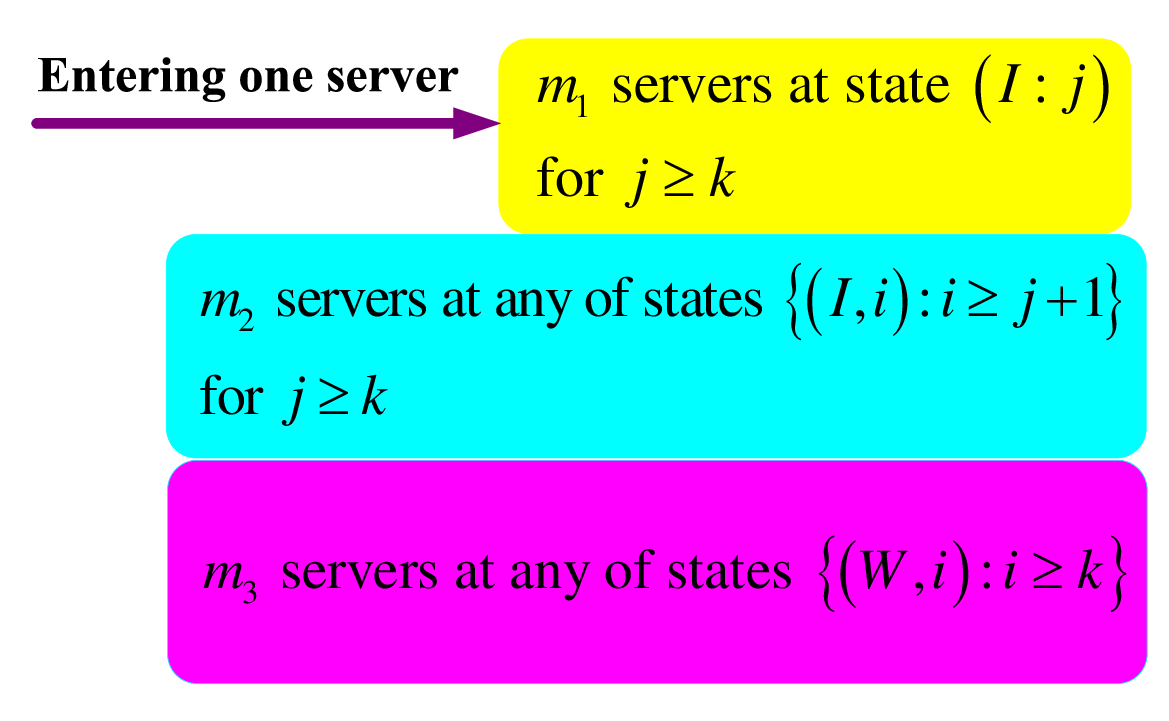} \caption{A primary
arriving customer enters one idle server}%
\label{figure: fig-3}%
\end{figure}

\textbf{Case three (Retrial arrivals and busy servers):} If the $d_{2}%
$\ chosen servers are all busy, a retrial arriving customer has to come back
to its original orbit again. In this case, the behavior of this retrial
supermarket model will not be changed, hence we should ignore this case.

\textbf{Case four (Retrial arrivals and idle servers):} If there exist some
idle servers among the $d_{2}$\ chosen servers and each orbit of the $d_{2}%
$\ chosen servers contains at least $k$ customers, a retrial arriving customer
enters one idle server whose orbit has the fewest $j$ customers for $j\geq
k+1$, and then receives service immediately. Clearly, each orbit of the other
idle servers contains at least $j$ customers, if any. In this case, the
corresponding rate is given by%
\begin{align}
&  \sum_{j=k+1}^{\infty}N\left(  j\theta\right)  \left[  u_{I,j}^{\left(
N\right)  }\left(  t\right)  -u_{I,j+1}^{\left(  N\right)  }\left(  t\right)
\right]  \sum_{m=0}^{d_{2}-1}C_{d_{2}-1}^{m}\left[  u_{I,j}^{\left(  N\right)
}\left(  t\right)  \right]  ^{m}\left[  u_{W,k}^{\left(  N\right)  }\left(
t\right)  \right]  ^{d_{2}-1-m}\text{d}t\nonumber\\
&  =N\theta\sum_{j=k+1}^{\infty}j\left[  u_{I,j}^{\left(  N\right)  }\left(
t\right)  -u_{I,j+1}^{\left(  N\right)  }\left(  t\right)  \right]  \left[
u_{I,j}^{\left(  N\right)  }\left(  t\right)  +u_{W,k}^{\left(  N\right)
}\left(  t\right)  \right]  ^{d_{2}-1}\text{d}t. \label{Equ-3}%
\end{align}

It is worthwhile to note that the condition: $j\geq k+1$, is necessary. This
can be explained from some state transitions of any retrial queue as follows:%
\[%
\begin{array}
[c]{c}%
\text{Server one: }\left(  I,j\right) \\
\downarrow j\theta\\
\text{Server two: }\left(  I,j\right)
\end{array}
\Longrightarrow%
\begin{array}
[c]{l}%
\text{Server one: }\left(  I,j-1\right)  \ \ \ j-1\geq k\text{ is necessary
for }u_{W,k}^{\left(  N\right)  }\left(  t\right) \\
\\
\text{Server two: }\left(  W,j\right)
\end{array}
\]

\textbf{Case five (Service processes):} If each orbit of the $N$ servers
contains at least $k$ customers, then the rate that any customer finishes its
required service and leaves this system is given by%
\begin{equation}
N\mu\left[  u_{W,k}^{\left(  N\right)  }\left(  t\right)  \right]  \text{d}t.
\label{Equ-4}%
\end{equation}

Based on the above analysis, it follows from (\ref{Equ-1}) to (\ref{Equ-4})
that%
\begin{align*}
\frac{\text{d}E\left[  n_{k}^{\left(  W\right)  }\left(  t\right)  \right]
}{\text{d}t} =  &  N\lambda\left[  \left(  u_{W,k-1}^{\left(  N\right)
}\left(  t\right)  \right)  ^{d_{1}}-\left(  u_{W,k}^{\left(  N\right)
}\left(  t\right)  \right)  ^{d_{1}}\right]  -N\mu\left[  u_{W,k}^{\left(
N\right)  }\left(  t\right)  \right] \\
&  +N\lambda\sum_{j=k}^{\infty}\left[  u_{I,j}^{\left(  N\right)  }\left(
t\right)  -u_{I,j+1}^{\left(  N\right)  }\left(  t\right)  \right]  \left[
u_{I,j}^{\left(  N\right)  }\left(  t\right)  +u_{W,k}^{\left(  N\right)
}\left(  t\right)  \right]  ^{d_{1}-1}\\
&  +N\theta\sum_{j=k+1}^{\infty}j\left[  u_{I,j}^{\left(  N\right)  }\left(
t\right)  -u_{I,j+1}^{\left(  N\right)  }\left(  t\right)  \right]  \left[
u_{I,j}^{\left(  N\right)  }\left(  t\right)  +u_{W,k}^{\left(  N\right)
}\left(  t\right)  \right]  ^{d_{2}-1},
\end{align*}
this, together with $u_{W,k}^{\left(  N\right)  }\left(  t\right)  =E\left[
n_{k}^{\left(  W\right)  }\left(  t\right)  /N\right]  $, gives%
\begin{align}
\frac{\text{d}u_{W,k}^{\left(  N\right)  }\left(  t\right)  }{\text{d}t} =  &
\lambda\left[  \left(  u_{W,k-1}^{\left(  N\right)  }\left(  t\right)
\right)  ^{d_{1}}-\left(  u_{W,k}^{\left(  N\right)  }\left(  t\right)
\right)  ^{d_{1}}\right]  -\mu u_{W,k}^{\left(  N\right)  }\left(  t\right)
\nonumber\\
&  +\lambda\sum_{j=k}^{\infty}\left[  u_{I,j}^{\left(  N\right)  }\left(
t\right)  -u_{I,j+1}^{\left(  N\right)  }\left(  t\right)  \right]  \left[
u_{I,j}^{\left(  N\right)  }\left(  t\right)  +u_{W,k}^{\left(  N\right)
}\left(  t\right)  \right]  ^{d_{1}-1}\nonumber\\
&  +\theta\sum_{j=k+1}^{\infty}j\left[  u_{I,j}^{\left(  N\right)  }\left(
t\right)  -u_{I,j+1}^{\left(  N\right)  }\left(  t\right)  \right]  \left[
u_{I,j}^{\left(  N\right)  }\left(  t\right)  +u_{W,k}^{\left(  N\right)
}\left(  t\right)  \right]  ^{d_{2}-1}. \label{Equ-5}%
\end{align}

Using a similar analysis to that derived in Equation (\ref{Equ-5}), we can
obtain a system of differential equations satisfied by the expected fraction
vector $\mathbf{u}_{N}\left(  t\right)  =(u_{0}^{\left(  N\right)  }\left(
t\right)  ,u_{1}^{\left(  N\right)  }\left(  t\right)  $, $u_{2}^{\left(
N\right)  }\left(  t\right)  ,\ldots)$ as follows: For $k\geq1$,%
\begin{align}
\frac{\text{d}u_{W,k}^{\left(  N\right)  }\left(  t\right)  }{\text{d}t}=  &
\lambda\left[  \left(  u_{W,k-1}^{\left(  N\right)  }\left(  t\right)
\right)  ^{d_{1}}-\left(  u_{W,k}^{\left(  N\right)  }\left(  t\right)
\right)  ^{d_{1}}\right]  -\mu u_{W,k}^{\left(  N\right)  }\left(  t\right)
\nonumber\\
&  +\lambda\sum_{j=k}^{\infty}\left[  u_{I,j}^{\left(  N\right)  }\left(
t\right)  -u_{I,j+1}^{\left(  N\right)  }\left(  t\right)  \right]  \left[
u_{I,j}^{\left(  N\right)  }\left(  t\right)  +u_{W,k}^{\left(  N\right)
}\left(  t\right)  \right]  ^{d_{1}-1}\nonumber\\
&  +\theta\sum_{j=k+1}^{\infty}j\left[  u_{I,j}^{\left(  N\right)  }\left(
t\right)  -u_{I,j+1}^{\left(  N\right)  }\left(  t\right)  \right]  \left[
u_{I,j}^{\left(  N\right)  }\left(  t\right)  +u_{W,k}^{\left(  N\right)
}\left(  t\right)  \right]  ^{d_{2}-1}, \label{Equat1}%
\end{align}
and for $l\geq0,$
\begin{align}
\frac{\text{d}u_{I,l}^{\left(  N\right)  }\left(  t\right)  }{\text{d}t}=  &
\mu u_{W,l}^{\left(  N\right)  }\left(  t\right)  -\lambda\sum_{j=l}^{\infty
}\left[  u_{I,j}^{\left(  N\right)  }\left(  t\right)  -u_{I,j+1}^{\left(
N\right)  }\left(  t\right)  \right]  \left[  u_{I,j}^{\left(  N\right)
}\left(  t\right)  +u_{W,l}^{\left(  N\right)  }\left(  t\right)  \right]
^{d_{1}-1}\nonumber\\
&  -\theta\sum_{j=l}^{\infty}j\left[  u_{I,j}^{\left(  N\right)  }\left(
t\right)  -u_{I,j+1}^{\left(  N\right)  }\left(  t\right)  \right]  \left[
u_{I,j}^{\left(  N\right)  }\left(  t\right)  +u_{W,l}^{\left(  N\right)
}\left(  t\right)  \right]  ^{d_{2}-1}, \label{Equat2}%
\end{align}
with the boundary condition%
\begin{equation}
u_{I,0}^{\left(  N\right)  }\left(  t\right)  +u_{W,0}^{\left(  N\right)
}\left(  t\right)  =1 \label{Equat3}%
\end{equation}
and the initial condition%
\begin{equation}
\left\{
\begin{array}
[c]{c}%
u_{W,k}^{\left(  N\right)  }\left(  t\right)  =g_{k},\\
u_{I,k}^{\left(  N\right)  }\left(  t\right)  =h_{k},
\end{array}
\right.  \text{ \ \ \ \ }k\geq0, \label{Equat4}%
\end{equation}
where%
\[
1\geq g_{0}\geq g_{1}\geq g_{2}\geq g_{3}\geq\cdots\geq0
\]
and%
\[
1\geq h_{0}\geq h_{1}\geq h_{2}\geq h_{3}\geq\cdots\geq0.
\]

To intuitively understand the system of differential equations (\ref{Equat1})
to (\ref{Equat4}), here it is necessary to derive the differential tail
equations in the retrial M/M/1 queue, that is, a special retrial supermarket
model with $d_{1}=d_{2}=1$. To that end, we denote by $\mathcal{Q}\left(
t\right)  $ and $C\left(  t\right)  $ the number of customers in the orbit and
the state of server at time $t$, where $\mathcal{Q}\left(  t\right)
=0,1,2,\ldots$ and $C\left(  t\right)  =W$ for the busy server or $I$ for the
idle server. For $k\geq0$, we write%
\[
p_{W,k}\left(  t\right)  =P\left\{  C\left(  t\right)  =W,\mathcal{Q}\left(
t\right)  =k\right\}
\]
and%
\[
p_{I,k}\left(  t\right)  =P\left\{  C\left(  t\right)  =I,\mathcal{Q}\left(
t\right)  =k\right\}  .
\]
From Figure 2, it is well-known that%
\begin{equation}
\left\{
\begin{array}
[c]{ll}%
\frac{\text{d}}{\text{d}t}p_{W,0}\left(  t\right)  =-\left(  \lambda
+\mu\right)  p_{W,0}\left(  t\right)  +\lambda p_{I,0}\left(  t\right)
+\theta p_{I,1}\left(  t\right)  , & k=0,\\
\frac{\text{d}}{\text{d}t}p_{W,k}\left(  t\right)  =-\left(  \lambda
+\mu\right)  p_{W,k}\left(  t\right)  +\lambda p_{I,k}\left(  t\right)
+\left(  k+1\right)  \theta p_{I,k+1}\left(  t\right)  +\lambda p_{W,k-1}%
\left(  t\right)  , & k\geq1,
\end{array}
\right.  \label{Equation1}%
\end{equation}%
\begin{equation}
\left\{
\begin{array}
[c]{ll}%
\frac{\text{d}}{\text{d}t}p_{I,0}\left(  t\right)  =\mu p_{W,0}\left(
t\right)  -\lambda p_{I,0}\left(  t\right)  , & k=0,\\
\frac{\text{d}}{\text{d}t}p_{I,k}\left(  t\right)  =\mu p_{W,k}\left(
t\right)  -\lambda p_{I,k}\left(  t\right)  -k\theta p_{I,k}\left(  t\right)
, & k\geq1.
\end{array}
\right.  \label{Equation2}%
\end{equation}

Let%
\[
\xi_{W,k}\left(  t\right)  =\sum_{l=k}^{\infty}p_{W,l}\left(  t\right)
,\text{ \ \ }\xi_{I,k}\left(  t\right)  =\sum_{l=k}^{\infty}p_{I,l}\left(
t\right)  .
\]
Note that%
\[
\sum_{j=k+1}^{\infty}jp_{I,j}\left(  t\right)  =\sum_{j=k+1}^{\infty}j\left[
\xi_{I,j}\left(  t\right)  -\xi_{I,j+1}\left(  t\right)  \right]  =\left(
k+1\right)  \xi_{I,k+1}\left(  t\right)  +\sum_{j=k+2}^{\infty}\xi
_{I,j}\left(  t\right)  ,
\]
it follows from (\ref{Equation1}) that for $k\geq1$%
\begin{align}
\frac{\text{d}}{\text{d}t}\xi_{W,k}\left(  t\right)  =  &  \lambda\left[
\xi_{W,k-1}\left(  t\right)  -\xi_{W,k}\left(  t\right)  \right]  -\mu
\xi_{W,k}\left(  t\right)  +\lambda\xi_{I,k}\left(  t\right) \nonumber\\
&  +\theta\left[  \left(  k+1\right)  \xi_{I,k+1}\left(  t\right)
+\sum_{j=k+2}^{\infty}\xi_{I,j}\left(  t\right)  \right]  . \label{Equation3}%
\end{align}
Similarly, it follows from (\ref{Equation2}) that for $k\geq0$%
\begin{equation}
\frac{\text{d}}{\text{d}t}\xi_{I,k}\left(  t\right)  =\mu\xi_{W,k}\left(
t\right)  -\lambda\xi_{I,k}\left(  t\right)  -\theta\left[  k\xi_{I,k}\left(
t\right)  +\sum_{j=k+1}^{\infty}\xi_{I,j}\left(  t\right)  \right]  .
\label{Equation4}%
\end{equation}
The boundary condition is given by%
\begin{equation}
\xi_{W,0}\left(  t\right)  +\xi_{I,0}\left(  t\right)  =1. \label{Equation5}%
\end{equation}
Therefore, it is easy to see that Equations (\ref{Equation3}) to
(\ref{Equation5}) are the same as Equations (\ref{Equat1}) to (\ref{Equat3})
for $d_{1}=d_{2}=1$.

\subsection{Useful upper bounds}

In this subsection, we provide two useful upper bounds for the number of
customers in any orbit of this retrial supermarket model, which will be
necessary and useful in our later study.

For $d_{1},d_{2}\geq1$, let $Q_{d_{1},d_{2}}^{\left(  N\right)  }\left(
t\right)  $, $Q_{d_{1},d_{2}}^{\left(  W,N\right)  }\left(  t\right)  $ and
$Q_{d_{1},d_{2}}^{\left(  I,N\right)  }\left(  t\right)  $ be the numbers of
customers in any orbit of this retrial supermarket model at time $t$ when the
corresponding server is at any state, busy and idle, respectively. Then%
\[
\mathcal{Q}\left(  t\right)  =Q_{1,1}^{\left(  N\right)  }\left(  t\right)
=Q_{1,1}^{\left(  W,N\right)  }\left(  t\right)  +Q_{1,1}^{\left(  I,N\right)
}\left(  t\right)  ,
\]
which is the queue length in the orbit of the M/M/1 retrial queue, then
$\mathcal{Q}\left(  t\right)  $ is independent of $N$.

Using a similar coupling method to Theorems 4 and 5 in Martin and Suhov
\cite{Mar:1999}, from the stochastic order we can obtain%
\begin{equation}
Q_{d_{1},d_{2}}^{\left(  N\right)  }\left(  t\right)  \leq\mathcal{Q}\left(
t\right)  , \label{Ineq-1}%
\end{equation}%
\begin{equation}
Q_{d_{1},d_{2}}^{\left(  W,N\right)  }\left(  t\right)  \leq Q_{1,1}^{\left(
W,N\right)  }\left(  t\right)  \leq\mathcal{Q}\left(  t\right)  \label{Ineq-2}%
\end{equation}
and%
\begin{equation}
Q_{d_{1},d_{2}}^{\left(  I,N\right)  }\left(  t\right)  \leq Q_{1,1}^{\left(
I,N\right)  }\left(  t\right)  \leq\mathcal{Q}\left(  t\right)  .
\label{Ineq-3}%
\end{equation}
Obviously, we also have%
\begin{equation}
\left[  Q_{d_{1},d_{2}}^{\left(  N\right)  }\left(  t\right)  \right]
^{2}\leq\left[  \mathcal{Q}\left(  t\right)  \right]  ^{2}, \label{Ineq-4}%
\end{equation}%
\begin{equation}
\left[  Q_{d_{1},d_{2}}^{\left(  W,N\right)  }\left(  t\right)  \right]
^{2}\leq\left[  Q_{1,1}^{\left(  W,N\right)  }\left(  t\right)  \right]
^{2}\leq\left[  \mathcal{Q}\left(  t\right)  \right]  ^{2} \label{Ineq-5}%
\end{equation}
and%
\begin{equation}
\left[  Q_{d_{1},d_{2}}^{\left(  I,N\right)  }\left(  t\right)  \right]
^{2}\leq\left[  Q_{1,1}^{\left(  I,N\right)  }\left(  t\right)  \right]
^{2}\leq\left[  \mathcal{Q}\left(  t\right)  \right]  ^{2}. \label{Ineq-6}%
\end{equation}

By using Excise one in Chapter 1 of Ross \cite{Ross:1983}, it is clear that
for $n\geq1$%
\[
E\left[  X^{n}\right]  =n\int_{0}^{+\infty}x^{n-1}\overline{F}\left(
x\right)  \text{d}x,
\]
or for a discrete random variable%
\[
E\left[  X^{n}\right]  =n\sum_{k=1}^{\infty}k^{n-1}\overline{P}_{k},
\]
where $\overline{P}_{k}=\sum_{j=k}^{\infty}p_{k}$ and $p_{k}=P\left\{
X=k\right\}  $. This gives%
\[
E\left[  Q_{d_{1},d_{2}}^{\left(  N\right)  }\left(  t\right)  \right]
=\sum_{k=1}^{\infty}\left[  u_{W,k}^{\left(  N\right)  }\left(  t\right)
+u_{I,k}^{\left(  N\right)  }\left(  t\right)  \right]  ,
\]%
\[
E\left[  \mathcal{Q}\left(  t\right)  \right]  =\sum_{k=1}^{\infty}\left[
\xi_{W,k}\left(  t\right)  +\xi_{I,k}\left(  t\right)  \right]
\]%
\[
E\left[  Q_{d_{1},d_{2}}^{\left(  W,N\right)  }\left(  t\right)  \right]
=\sum_{k=1}^{\infty}u_{W,k}^{\left(  N\right)  }\left(  t\right)  ,\text{
}E\left[  Q_{1,1}^{\left(  W,N\right)  }\left(  t\right)  \right]  =\sum
_{k=1}^{\infty}\xi_{W,k}\left(  t\right)  ,
\]%
\[
E\left[  Q_{d_{1},d_{2}}^{\left(  I,N\right)  }\left(  t\right)  \right]
=\sum_{k=1}^{\infty}u_{I,k}^{\left(  N\right)  }\left(  t\right)  ,\text{
}E\left[  Q_{1,1}^{\left(  I,N\right)  }\left(  t\right)  \right]  =\sum
_{k=1}^{\infty}\xi_{I,k}\left(  t\right)  ;
\]%
\[
\sum_{k=1}^{\infty}k\left[  u_{W,k}^{\left(  N\right)  }\left(  t\right)
+u_{I,k}^{\left(  N\right)  }\left(  t\right)  \right]  =\frac{1}{2}E\left[
\left[  Q_{d_{1},d_{2}}^{\left(  N\right)  }\left(  t\right)  \right]
^{2}\right]  ,
\]%
\[
\sum_{k=1}^{\infty}k\left[  \xi_{W,k}\left(  t\right)  +\xi_{I,k}\left(
t\right)  \right]  =\frac{1}{2}E\left[  \left[  \mathcal{Q}\left(  t\right)
\right]  ^{2}\right]  ,
\]%
\[
\sum_{k=1}^{\infty}ku_{W,k}^{\left(  N\right)  }\left(  t\right)  =\frac{1}%
{2}E\left[  \left[  Q_{d_{1},d_{2}}^{\left(  W,N\right)  }\left(  t\right)
\right]  ^{2}\right]  ,\text{ }\sum_{k=1}^{\infty}k\xi_{W,k}\left(  t\right)
=\frac{1}{2}E\left[  \left[  Q_{1,1}^{\left(  W,N\right)  }\left(  t\right)
\right]  ^{2}\right]  ,
\]
and%
\[
\sum_{k=1}^{\infty}ku_{I,k}^{\left(  N\right)  }\left(  t\right)  =\frac{1}%
{2}E\left[  \left[  Q_{d_{1},d_{2}}^{\left(  I,N\right)  }\left(  t\right)
\right]  ^{2}\right]  ,\text{ }\sum_{k=1}^{\infty}k\xi_{I,k}\left(  t\right)
=\frac{1}{2}E\left[  \left[  Q_{1,1}^{\left(  I,N\right)  }\left(  t\right)
\right]  ^{2}\right]  .
\]
Thus, it follows from (\ref{Ineq-1}) to (\ref{Ineq-6}) that%
\begin{equation}
\sum_{k=1}^{\infty}\left[  u_{W,k}^{\left(  N\right)  }\left(  t\right)
+u_{I,k}^{\left(  N\right)  }\left(  t\right)  \right]  \leq E\left[
\mathcal{Q}\left(  t\right)  \right]  , \label{Ineq-7-1}%
\end{equation}%
\begin{equation}
\sum_{k=1}^{\infty}u_{W,k}^{\left(  N\right)  }\left(  t\right)  \leq E\left[
\mathcal{Q}\left(  t\right)  \right]  , \label{Ineq-7}%
\end{equation}%
\begin{equation}
\sum_{k=1}^{\infty}u_{I,k}^{\left(  N\right)  }\left(  t\right)  \leq E\left[
\mathcal{Q}\left(  t\right)  \right]  ; \label{Ineq-8}%
\end{equation}%
\begin{equation}
\sum_{k=1}^{\infty}k\left[  u_{W,k}^{\left(  N\right)  }\left(  t\right)
+u_{I,k}^{\left(  N\right)  }\left(  t\right)  \right]  \leq\frac{1}%
{2}E\left[  \mathcal{Q}\left(  t\right)  ^{2}\right]  , \label{Ineq-9-1}%
\end{equation}%
\begin{equation}
\sum_{k=1}^{\infty}ku_{W,k}^{\left(  N\right)  }\left(  t\right)  \leq\frac
{1}{2}E\left[  \mathcal{Q}\left(  t\right)  ^{2}\right]  \label{Ineq-9}%
\end{equation}
and%
\begin{equation}
\sum_{k=1}^{\infty}ku_{I,k}^{\left(  N\right)  }\left(  t\right)  \leq\frac
{1}{2}E\left[  \mathcal{Q}\left(  t\right)  ^{2}\right]  . \label{Ineq-10}%
\end{equation}

The following theorem provides useful upper bounds for the series $\sum
_{k=1}^{\infty}u_{W,k}^{\left(  N\right)  }\left(  t\right)  $, $\sum
_{k=1}^{\infty}u_{I,k}^{\left(  N\right)  }\left(  t\right)  $, $\sum
_{k=1}^{\infty}ku_{W,k}^{\left(  N\right)  }\left(  t\right)  $ and
$\sum_{k=1}^{\infty}ku_{I,k}^{\left(  N\right)  }\left(  t\right)  $, which
will be necessary and useful in our later study.

\begin{The}
\label{The:Ineq}For any $t\geq0$ and $N=1,2,3,\ldots$, if $\rho<1$, then there
exists two bigger numbers $C_{1},C_{2}>0$ such that%
\begin{equation}
\max\left\{  \sum_{k=1}^{\infty}u_{W,k}^{\left(  N\right)  }\left(  t\right)
,\sum_{k=1}^{\infty}u_{I,k}^{\left(  N\right)  }\left(  t\right)  \right\}
\leq\sum_{k=1}^{\infty}\left[  u_{W,k}^{\left(  N\right)  }\left(  t\right)
+u_{I,k}^{\left(  N\right)  }\left(  t\right)  \right]  <C_{1} \label{Ineq-11}%
\end{equation}
and%
\begin{equation}
\max\left\{  \sum_{k=1}^{\infty}ku_{W,k}^{\left(  N\right)  }\left(  t\right)
,\sum_{k=1}^{\infty}ku_{I,k}^{\left(  N\right)  }\left(  t\right)  \right\}
\leq\sum_{k=1}^{\infty}k\left[  u_{W,k}^{\left(  N\right)  }\left(  t\right)
+u_{I,k}^{\left(  N\right)  }\left(  t\right)  \right]  <C_{2}.
\label{Ineq-12}%
\end{equation}

\end{The}

\textbf{Proof:} \ We only prove (\ref{Ineq-11}), while (\ref{Ineq-12}) can be
proved similarly by means of (2.14) and (2.15) in Kulkarni and Liang
\cite{Kul:1997}.

If $\rho<1$, then the M/M/1 retrial queue is stable. It follows from (2.14) in
Kulkarni and Liang \cite{Kul:1997} that%
\[
\lim_{t\rightarrow+\infty}E\left[  \mathcal{Q}\left(  t\right)  \right]
=\frac{\lambda\rho}{1-\rho}\left(  \frac{1}{\mu}+\frac{1}{\theta}\right)  .
\]
Hence, for an arbitrarily small $\varepsilon>0$, there exists a sufficiently
big $T>0$ such that for $t>T$%
\begin{equation}
E\left[  \mathcal{Q}\left(  t\right)  \right]  <\frac{\lambda\rho}{1-\rho
}\left(  \frac{1}{\mu}+\frac{1}{\theta}\right)  +\varepsilon. \label{Ineq-13}%
\end{equation}
Since $E\left[  \mathcal{Q}\left(  t\right)  \right]  $ is a continuous
function for $t\in\left[  0,T\right]  $, there exists a bigger number
$D_{1}>0$ such that%
\begin{equation}
\max_{t\in\left[  0,T\right]  }\left\{  E\left[  \mathcal{Q}\left(  t\right)
\right]  \right\}  \leq D_{1}. \label{Ineq-14}%
\end{equation}
It follows from (\ref{Ineq-13}) and (\ref{Ineq-14}) that%
\[
\sup_{t\geq0}\left\{  E\left[  \mathcal{Q}\left(  t\right)  \right]  \right\}
\leq\max\left\{  D_{1},\frac{\lambda\rho}{1-\rho}\left(  \frac{1}{\mu}%
+\frac{1}{\theta}\right)  +\varepsilon\right\}  =C_{1}.
\]
It is easy to see from (\ref{Ineq-7-1}) to (\ref{Ineq-8}) that (\ref{Ineq-11})
always hold. This completes the proof. \textbf{{\rule{0.08in}{0.08in}}}

\section{Solution to the System of ODEs}

Throughout this section, we assume that this limit: $\mathbf{u}(t)=\lim
_{N\rightarrow\infty}\mathbf{u}^{(N)}(t)$ always exists and $\mathbf{u}%
(t)\gneqq0$ for $t\geq0$ (note that the next section will prove the existence
of this limit by means of the mean-field theory). In this case, we write that
$u_{W,k}\left(  t\right)  =\lim_{N\rightarrow\infty}u_{W,k}^{(N)}(t)$ and
$u_{I,k}\left(  t\right)  =\lim_{N\rightarrow\infty}u_{I,k}^{(N)}(t)$ for
$k\geq0$ and $t\geq0$. Therefore, it follows from the system of differential
equations (\ref{Equat1}) to (\ref{Equat4}) as $N\rightarrow\infty$ that
$\mathbf{u}(t)=\left(  u_{W,0}\left(  t\right)  ,u_{I,0}\left(  t\right)
,u_{W,1}\left(  t\right)  ,u_{I,1}\left(  t\right)  ,u_{W,2}\left(  t\right)
,u_{I,2}\left(  t\right)  ,\ldots\right)  $ is a solution to the following
system of limiting differential equations: For $k\geq1$%
\begin{align}
\frac{\text{d}u_{W,k}\left(  t\right)  }{\text{d}t}=  &  \lambda\left[
\left(  u_{W,k-1}\left(  t\right)  \right)  ^{d_{1}}-\left(  u_{W,k}\left(
t\right)  \right)  ^{d_{1}}\right]  -\mu u_{W,k}\left(  t\right) \nonumber\\
&  +\lambda\sum_{j=k}^{\infty}\left[  u_{I,j}\left(  t\right)  -u_{I,j+1}%
\left(  t\right)  \right]  \left[  u_{I,j}\left(  t\right)  +u_{W,k}\left(
t\right)  \right]  ^{d_{1}-1}\nonumber\\
&  +\theta\sum_{j=k+1}^{\infty}j\left[  u_{I,j}\left(  t\right)
-u_{I,j+1}\left(  t\right)  \right]  \left[  u_{I,j}\left(  t\right)
+u_{W,k}\left(  t\right)  \right]  ^{d_{2}-1}, \label{Equat13-0}%
\end{align}
and for $l\geq0$
\begin{align}
\frac{\text{d}u_{I,l}\left(  t\right)  }{\text{d}t}=  &  \mu u_{W,l}\left(
t\right)  -\lambda\sum_{j=l}^{\infty}\left[  u_{I,j}\left(  t\right)
-u_{I,j+1}\left(  t\right)  \right]  \left[  u_{I,j}\left(  t\right)
+u_{W,l}\left(  t\right)  \right]  ^{d_{1}-1}\nonumber\\
&  -\theta\sum_{j=l}^{\infty}j\left[  u_{I,j}\left(  t\right)  -u_{I,j+1}%
\left(  t\right)  \right]  \left[  u_{I,j}\left(  t\right)  +u_{W,l}\left(
t\right)  \right]  ^{d_{2}-1}, \label{Equat14}%
\end{align}
with the boundary condition%
\begin{equation}
u_{I,0}\left(  t\right)  +u_{W,0}\left(  t\right)  =1, \label{Equat16}%
\end{equation}
and the initial conditions%
\begin{equation}
\left\{
\begin{array}
[c]{c}%
u_{W,k}\left(  0\right)  =g_{k},\\
u_{I,k}\left(  0\right)  =h_{k},
\end{array}
\right.  \text{ \ \ }k\geq0. \label{Equat17}%
\end{equation}

Let $\mathbf{u}(t)=\mathbf{u}(t,\mathbf{g},\mathbf{h})$ be a solution to the
system of differential equations (\ref{Equat13-0}) to (\ref{Equat17}) for
$t\geq0$. Then $\mathbf{u}(0)=\mathbf{u}(0,\mathbf{g},\mathbf{h})=\left(
\mathbf{g},\mathbf{h}\right)  $, where $\mathbf{g}=\left(  g_{0},g_{1}%
,g_{2},\ldots\right)  \ $and $\mathbf{h}=\left(  h_{0},h_{1},h_{2}%
,\ldots\right)  $.

It follows from (\ref{Equat0}) that as $N\rightarrow\infty$%
\[
1\geq u_{W,0}\left(  t\right)  \geq u_{W,1}\left(  t\right)  \geq
u_{W,2}\left(  t\right)  \geq\cdots\geq0
\]
and%
\[
1\geq u_{I,0}\left(  t\right)  \geq u_{I,1}\left(  t\right)  \geq
u_{I,2}\left(  t\right)  \geq\cdots\geq0.
\]

\vskip0.3cm

Now, we prove the existence and uniqueness of the solution to the system of
differential equations (\ref{Equat13-0}) to (\ref{Equat17}). To that end, we
need to provide a computational method for establishing a Lipschitz condition
for $d_{1}\geq1$ and $d_{2}\geq2$, which is a more difficult issue in the
literature of supermarket models. It is worthwhile to note that once the
Lipschitz condition is obtained, the proof of the existence and uniqueness of
the solution can be given easily through the Picard approximation as well as
some basic results of the Banach space, e.g., see Li et al \cite{Li:2013} for
more details. Note that the retrial discipline makes computation of such a
Lipschitz condition more complicated, as seen in our later analysis.

To provide the Lipschitz condition, we need to use the derivative of the
infinite-dimensional vector function $G:\mathbf{R}_{+}^{\infty}\rightarrow
\mathbf{C}^{1}\left(  \mathbf{R}_{+}^{\infty}\right)  $. Here, for the
convenience of description, we restate some useful notation, definitions and
results given in Section 4.1 of Li et al \cite{Li:2013}.

For the infinite-dimensional vector function $G:\mathbf{R}_{+}^{\infty
}\rightarrow\mathbf{C}^{1}\left(  \mathbf{R}_{+}^{\infty}\right)  $, we write
$x=(x_{1},x_{2},x_{3},\ldots)$ and $G(x)=(G_{1}(x),G_{2}(x),G_{3}(x),\ldots)$,
where $x_{k}$ and $G_{k}(x)$ are scalar for $k\geq1$. Then the matrix of
partial derivatives of the infinite-dimensional vector function $G(x)$ is
defined as%
\begin{equation}
\mathcal{D}G(x)=\left(
\begin{array}
[c]{cccc}%
\dfrac{\partial G_{1}(x)}{\partial x_{1}} & \dfrac{\partial G_{2}(x)}{\partial
x_{1}} & \dfrac{\partial G_{3}(x)}{\partial x_{1}} & \cdots\\
\dfrac{\partial G_{1}(x)}{\partial x_{2}} & \dfrac{\partial G_{2}(x)}{\partial
x_{2}} & \dfrac{\partial G_{3}(x)}{\partial x_{2}} & \cdots\\
\dfrac{\partial G_{1}(x)}{\partial x_{3}} & \dfrac{\partial G_{2}(x)}{\partial
x_{3}} & \dfrac{\partial G_{3}(x)}{\partial x_{3}} & \cdots\\
\vdots & \vdots & \vdots &
\end{array}
\right)  , \label{Eq7.3}%
\end{equation}
if each of the above partial derivatives exists.

For the infinite-dimensional vector function $G:\mathbf{R}_{+}^{\infty
}\rightarrow\mathbf{C}^{1}\left(  \mathbf{R}_{+}^{\infty}\right)  $, if there
exists a linear operator $A:\mathbf{R}_{+}^{\infty}\rightarrow\mathbf{C}%
^{1}\left(  \mathbf{R}_{+}^{\infty}\right)  $ such that for any vector
$\mathbf{f}\in\mathbf{R}^{\infty}$ and a non-zero scalar $t\in\mathbf{R}$%
\[
\lim_{t\rightarrow0}\frac{||G\left(  x+t\mathbf{f}\right)  -G\left(  x\right)
-t\mathbf{f}A||}{t}=0,
\]
then the function $G\left(  x\right)  $ is called to be Gateaux differentiable
at $x\in\mathbf{R}_{+}^{\infty}$. In this case, we write the Gateaux
derivative $G_{G}^{\prime}(x)=A$. In fact, $G_{G}^{\prime}(x)=\mathcal{D}G(x)$.

Let $\boldsymbol{t} =\left(  t_{1},t_{2},t_{3},\ldots\right)  $ with $0\leq
t_{k}\leq1$ for $k\geq1$. Then we write%
\[
\mathcal{D}G(x+\boldsymbol{t\oslash} \left(  y-x\right)  )=\left(
\begin{array}
[c]{cccc}%
\dfrac{\partial G_{1}(x+t_{1}\left(  y-x\right)  )}{\partial x_{1}} &
\dfrac{\partial G_{2}(x+t_{2}\left(  y-x\right)  )}{\partial x_{1}} &
\dfrac{\partial G_{3}(x+t_{3}\left(  y-x\right)  )}{\partial x_{1}} & \cdots\\
\dfrac{\partial G_{1}(x+t_{1}\left(  y-x\right)  )}{\partial x_{2}} &
\dfrac{\partial G_{2}(x+t_{2}\left(  y-x\right)  )}{\partial x_{2}} &
\dfrac{\partial G_{3}(x+t_{3}\left(  y-x\right)  )}{\partial x_{2}} & \cdots\\
\dfrac{\partial G_{1}(x+t_{1}\left(  y-x\right)  )}{\partial x_{3}} &
\dfrac{\partial G_{2}(x+t_{2}\left(  y-x\right)  )}{\partial x_{3}} &
\dfrac{\partial G_{3}(x+t_{3}\left(  y-x\right)  )}{\partial x_{3}} & \cdots\\
\vdots & \vdots & \vdots &
\end{array}
\right)  .
\]

If the infinite-dimensional vector function $G:\mathbf{R}_{+}^{\infty
}\rightarrow\mathbf{C}^{1}\left(  \mathbf{R}_{+}^{\infty}\right)  $ is Gateaux
differentiable, then there exists a vector $\boldsymbol{t}=\left(  t_{1}%
,t_{2},t_{3},\ldots\right)  $ with $0\leq t_{k}\leq1$ for $k\geq1$ such that%
\begin{equation}
G\left(  y\right)  -G\left(  x\right)  =\left(  y-x\right)  \mathcal{D}%
G(x+\boldsymbol{t\oslash}\left(  y-x\right)  ). \label{Eq7.3.2}%
\end{equation}
Furthermore, we have%
\begin{equation}
||G\left(  y\right)  -G\left(  x\right)  ||\text{ }\leq\sup_{0\leq t\leq
1}||\mathcal{D}G(x+t\left(  y-x\right)  )||\text{ }||y-x||. \label{Eq7.3.3}%
\end{equation}

Let $x=\left(  x_{0},x_{1},x_{2},\ldots\right)  $ with $x_{k}=\left(
x_{I,k},x_{W,k}\right)  =\left(  u_{I,k}\left(  t\right)  ,u_{W,k}\left(
t\right)  \right)  $ for $k\geq0$. Similarly, set $F(x)=(F_{0}(x),F_{1}%
(x),F_{2}(x),\ldots)$ with $F_{k}(x)=\left(  F_{I,k}(x),F_{W,k}(x)\right)  $
for $k\geq0$.

From (\ref{Equat13-0}) to (\ref{Equat17}), we write that for $k\geq1$
\begin{align}
F_{W,k}(x)=  &  \lambda\left(  x_{W,k-1}^{d_{1}}-x_{W,k}^{d_{1}}\right)  -\mu
x_{W,k}+\lambda\sum_{j=k}^{\infty}\left(  x_{I,j}-x_{I,j+1}\right)  \left(
x_{I,j}+x_{W,k}\right)  ^{d_{1}-1}\nonumber\\
&  +\theta\sum_{j=k+1}^{\infty}j\left(  x_{I,j}-x_{I,j+1}\right)  \left(
x_{I,j}+x_{W,k}\right)  ^{d_{2}-1}, \label{Equ5.5}%
\end{align}
and for $l\geq0$%
\begin{align}
F_{I,l}(x)=  &  \mu x_{W,l}-\lambda\sum_{j=l}^{\infty}\left(  x_{I,j}%
-x_{I,j+1}\right)  \left(  x_{I,j}+x_{W,l}\right)  ^{d_{1}-1}\nonumber\\
&  -\theta\sum_{j=l}^{\infty}j\left(  x_{I,j}-x_{I,j+1}\right)  \left(
x_{I,j}+x_{W,l}\right)  ^{d_{2}-1}, \label{Equ5.6}%
\end{align}
with%
\begin{equation}
x_{W,0}+x_{I,0}=1. \label{Equ5.7}%
\end{equation}
It is clear that $F(x)$ is in $\mathbf{C}^{2}\left(  \mathbf{R}_{+}^{\infty
}\right)  $. At the same time, it is easy to see that the system of
differential equations (\ref{Equat13-0}) to (\ref{Equat17}) is written as%
\begin{equation}
\frac{\text{d}}{\text{d}t}x=F(x) \label{Equ5.8}%
\end{equation}
with the initial condition%
\begin{equation}
x\left(  0\right)  =\left(  \mathbf{g},\mathbf{h}\right)  . \label{Equ5.9}%
\end{equation}

In what follows we will show that the infinite-dimensional vector function
$F(x)$ is Lipschitz. To that end, it is easy to see from (\ref{Equ5.5}) to
(\ref{Equ5.7}) that the matrix of partial derivatives of the function $F(x)$
is given by%
\begin{equation}
\mathcal{D}F(x)=\left(
\begin{array}
[c]{ccccc}%
A_{0,0}\left(  x\right)  & A_{0,1}\left(  x\right)  &  &  & \\
A_{1,0}\left(  x\right)  & A_{1,1}\left(  x\right)  & A_{1,2}\left(  x\right)
&  & \\
A_{2,0}\left(  x\right)  & A_{2,1}\left(  x\right)  & A_{2,2}\left(  x\right)
& A_{2,3}\left(  x\right)  & \\
\vdots & \vdots & \vdots & \vdots & \ddots
\end{array}
\right)  , \label{Equ5.10}%
\end{equation}
where for $i,j\geq0$%
\[
A_{i,j}\left(  x\right)  =\left(
\begin{array}
[c]{cc}%
\dfrac{\text{d}F_{I,j}(x)}{\partial x_{I,i}} & \dfrac{\text{d}F_{W,j}%
(x)}{\partial x_{I,i}}\\
\dfrac{\text{d}F_{I,j}(x)}{\partial x_{W,i}} & \dfrac{\text{d}F_{W,j}%
(x)}{\partial x_{W,i}}%
\end{array}
\right)  ,
\]

Now, using (\ref{Equ5.5}) to (\ref{Equ5.7}) we compute the matrices
$A_{i,j}\left(  x\right)  $ for $i,j\geq0$. To this end, for $k\geq0$ we write%
\[
L\left(  k\right)  =\sum_{j=k}^{\infty}\left(  x_{I,j}-x_{I,j+1}\right)
\left(  x_{I,j}+x_{W,k}\right)  ^{d_{1}-1}%
\]
and%
\[
M\left(  k\right)  =\sum_{j=k}^{\infty}j\left(  x_{I,j}-x_{I,j+1}\right)
\left(  x_{I,j}+x_{W,k}\right)  ^{d_{2}-1}.
\]
Then for $0\leq k\leq i-1$ and $i\geq1$%
\begin{align}
\dfrac{\text{d}L(k)}{\text{d}x_{I,i}}=  &  \left(  x_{I,i}+x_{W,k}\right)
^{d_{1}-1}-\left(  x_{I,i-1}+x_{W,k}\right)  ^{d_{1}-1}\nonumber\\
&  +\left(  d_{1}-1\right)  \left(  x_{I,i}-x_{I,i+1}\right)  \left(
x_{I,i}+x_{W,k}\right)  ^{d_{1}-2}, \label{In-1}%
\end{align}%
\begin{align}
\dfrac{\text{d}M(k)}{\text{d}x_{I,i}}=  &  i\left(  x_{I,i}+x_{W,k}\right)
^{d_{2}-1}-\left(  i-1\right)  \left(  x_{I,i-1}+x_{W,k}\right)  ^{d_{2}%
-1}\nonumber\\
&  +\left(  d_{2}-1\right)  i\left(  x_{I,i}-x_{I,i+1}\right)  \left(
x_{I,i}+x_{W,k}\right)  ^{d_{2}-2}, \label{In-2}%
\end{align}%
\[
\dfrac{\text{d}L(k)}{\text{d}x_{W,i}}=0,
\]%
\[
\dfrac{\text{d}M(k)}{\text{d}x_{W,i}}=0;
\]
for $k=i\geq0$%
\begin{equation}
\frac{\text{d}L\left(  i\right)  }{\text{d}x_{I,i}}=\left(  x_{I,i}%
+x_{W,i}\right)  ^{d_{1}-1}+\left(  d_{1}-1\right)  \left(  x_{I,i}%
-x_{I,i+1}\right)  \left(  x_{I,i}+x_{W,i}\right)  ^{d_{1}-2}, \label{In-3}%
\end{equation}%
\begin{equation}
\frac{\text{d}M\left(  i\right)  }{\text{d}x_{I,i}}=i\left(  x_{I,i}%
+x_{W,i}\right)  ^{d_{2}-1}+\left(  d_{2}-1\right)  i\left(  x_{I,i}%
-x_{I,i+1}\right)  \left(  x_{I,i}+x_{W,i}\right)  ^{d_{1}-2}, \label{In-4}%
\end{equation}%
\begin{equation}
\frac{\text{d}L\left(  i\right)  }{\text{d}x_{W,i}}=\left(  d_{1}-1\right)
\sum_{j=i}^{\infty}\left(  x_{I,j}-x_{I,j+1}\right)  \left(  x_{I,j}%
+x_{W,i}\right)  ^{d_{1}-2}, \label{In-5}%
\end{equation}%
\begin{equation}
\frac{\text{d}M\left(  i\right)  }{\text{d}x_{W,i}}=\left(  d_{2}-1\right)
\sum_{j=i}^{\infty}j\left(  x_{I,j}-x_{I,j+1}\right)  \left(  x_{I,j}%
+x_{W,i}\right)  ^{d_{2}-2}. \label{In-6}%
\end{equation}
Specifically, we have%
\[
\frac{\text{d}M\left(  i+1\right)  }{\text{d}x_{I,i}}=0,
\]%
\[
\frac{\text{d}M\left(  i+1\right)  }{\text{d}x_{W,i}}=0;
\]
for $k=i-1$%
\[
M\left(  k+1\right)  =M\left(  i\right)  ,
\]
and for $0\leq k\leq i-2$%
\begin{align*}
\dfrac{\text{d}M(k+1)}{\text{d}x_{I,i}}=  &  i\left(  x_{I,i}+x_{W,k+1}%
\right)  ^{d_{2}-1}-\left(  i-1\right)  \left(  x_{I,i-1}+x_{W,k+1}\right)
^{d_{2}-1}\\
&  +i\left(  d_{2}-1\right)  \left(  x_{I,i}-x_{I,i+1}\right)  \left(
x_{I,i}+x_{W,k+1}\right)  ^{d_{1}-2}%
\end{align*}
and%
\[
\dfrac{\text{d}M(k+1)}{\text{d}x_{W,i}}=0.
\]

From (\ref{Equ5.7}), it is clear that%
\[
\frac{\text{d}}{\text{d}t}x_{W,0}+\frac{\text{d}}{\text{d}t}x_{I,0}=0,
\]
this gives%
\[
F_{W,0}(x)=-F_{I,0}(x).
\]
Hence we obtain%
\[
A_{0,0}(x)=\left(
\begin{array}
[c]{cc}%
-\lambda\dfrac{\text{d}L(0)}{\text{d}x_{I,0}} & \lambda\dfrac{\text{d}%
L(0)}{\text{d}x_{I,0}}\\
\mu-\lambda\dfrac{\text{d}L(0)}{\text{d}x_{W,0}}-\theta\dfrac{\text{d}%
M(0)}{\text{d}x_{W,0}} & -\mu+\lambda\dfrac{\text{d}L(0)}{\text{d}x_{W,0}%
}+\theta\dfrac{\text{d}M(0)}{\text{d}x_{W,0}}%
\end{array}
\right)
\]
and%
\[
A_{0,1}(x)=\left(
\begin{array}
[c]{cc}%
0 & 0\\
0 & \lambda d_{1}x_{W,0}^{d_{1}-1}%
\end{array}
\right)  ;
\]
for $0\leq k\leq i-1$ and $i\geq1$%
\begin{align}
A_{i,k}\left(  x\right)   &  =\left(
\begin{array}
[c]{cc}%
-\lambda\dfrac{\text{d}L(k)}{\text{d}x_{I,i}}-\theta\dfrac{\text{d}%
M(k)}{\text{d}x_{I,i}} & \lambda\dfrac{\text{d}L(k)}{\text{d}x_{I,i}}%
+\theta\dfrac{\text{d}M(k+1)}{\text{d}x_{I,i}}\\
-\lambda\dfrac{\text{d}L(k)}{\text{d}x_{W,i}}-\theta\dfrac{\text{d}%
M(k)}{\text{d}x_{W,i}} & \lambda\dfrac{\text{d}L(k)}{\text{d}x_{W,i}}%
+\theta\dfrac{\text{d}M(k+1)}{\text{d}x_{W,i}}%
\end{array}
\right) \nonumber\\
&  =\left(
\begin{array}
[c]{cc}%
-\lambda\dfrac{\text{d}L(k)}{\text{d}x_{I,i}}-\theta\dfrac{\text{d}%
M(k)}{\text{d}x_{I,i}} & \lambda\dfrac{\text{d}L(k)}{\text{d}x_{I,i}}%
+\theta\dfrac{\text{d}M(k+1)}{\text{d}x_{I,i}}\\
0 & \theta\dfrac{\text{d}M(k+1)}{\text{d}x_{W,i}}%
\end{array}
\right)  \label{A-1}%
\end{align}
with%
\[
\dfrac{\text{d}M(k+1)}{\text{d}x_{W,i}}=\left\{
\begin{array}
[c]{cc}%
0, & 0\leq k\leq i-2,\\
\dfrac{\text{d}M(i)}{\text{d}x_{W,i}}, & k=i-1,
\end{array}
\right.
\]
for $k=i\geq1$
\begin{align}
A_{i,i}\left(  x\right)   &  =\left(
\begin{array}
[c]{cc}%
-\lambda\dfrac{\text{d}L(i)}{\text{d}x_{I,i}}-\theta\dfrac{\text{d}%
M(i)}{\text{d}x_{I,i}} & \lambda\dfrac{\text{d}L(i)}{\text{d}x_{I,i}}%
+\theta\dfrac{\text{d}M(i+1)}{\text{d}x_{I,i}}\\
\mu-\lambda\dfrac{\text{d}L(i)}{\text{d}x_{W,i}}-\theta\dfrac{\text{d}%
M(i)}{\text{d}x_{W,i}} & -\mu-\lambda d_{1}x_{W,i}^{d_{1}-1}+\lambda
\dfrac{\text{d}L(i)}{\text{d}x_{W,i}}+\theta\dfrac{\text{d}M(i+1)}%
{\text{d}x_{W,i}}%
\end{array}
\right) \nonumber\\
&  =\left(
\begin{array}
[c]{cc}%
-\lambda\dfrac{\text{d}L(i)}{\text{d}x_{I,i}}-\theta\dfrac{\text{d}%
M(i)}{\text{d}x_{I,i}} & \lambda\dfrac{\text{d}L(i)}{\text{d}x_{I,i}}\\
\mu-\lambda\dfrac{\text{d}L(i)}{\text{d}x_{W,i}}-\theta\dfrac{\text{d}%
M(i)}{\text{d}x_{W,i}} & -\lambda d_{1}x_{W,i}^{d_{1}-1}-\mu+\lambda
\dfrac{\text{d}L(i)}{\text{d}x_{W,i}}%
\end{array}
\right)  , \label{A-2}%
\end{align}
and for $k=i+1\geq2$%
\begin{equation}
A_{i,i+1}\left(  x\right)  =\left(
\begin{array}
[c]{cc}%
0 & 0\\
0 & \lambda d_{1}x_{W,i}^{d_{1}-1}%
\end{array}
\right)  . \label{A-3}%
\end{equation}

Now, we set up some useful bounds for the vector $x=\left(  x_{0},x_{1}%
,x_{2},\ldots\right)  $ with $x_{k}=\left(  x_{I,k},x_{W,k}\right)  $. Let%
\[
x_{W,k}=\lim_{N\rightarrow\infty}u_{W,k}^{\left(  N\right)  }\left(  t\right)
\]
and%
\[
x_{I,k}=\lim_{N\rightarrow\infty}u_{I,k}^{\left(  N\right)  }\left(  t\right)
.
\]
Then it follows from Theorem \ref{The:Ineq} that for any $t\geq0$, if $\rho
<1$, then there exists two bigger numbers $C_{1},C_{2}>0$ such that%
\begin{equation}
\max\left\{  \sum_{k=1}^{\infty}x_{W,k},\sum_{k=1}^{\infty}x_{I,k}\right\}
\leq\sum_{k=1}^{\infty}\left[  x_{W,k}+x_{I,k}\right]  <C_{1} \label{Ineq-15}%
\end{equation}
and%
\begin{equation}
\max\left\{  \sum_{k=1}^{\infty}kx_{W,k},\sum_{k=1}^{\infty}kx_{I,k}\right\}
\leq\sum_{k=1}^{\infty}k\left[  x_{W,k}+x_{I,k}\right]  <C_{2}.
\label{Ineq-16}%
\end{equation}

The following theorem provides two useful inequalities, both of which will be
necessary in establishing the bound of $\left\Vert \mathcal{D}F(x)\right\Vert
$.

\begin{The}
\label{The:Upper}If $\rho<1$, then%
\begin{equation}
\sup_{i\geq0}\left\{  \sum\limits_{k=0}^{i}\left\vert \dfrac{\text{d}%
L(k)}{\text{d}x_{I,i}}\right\vert \right\}  \leq1+\left(  d_{1}-1\right)
\left(  3+2C_{1}\right)  \label{Inequ-1}%
\end{equation}
and for $d_{2}\geq2$%
\begin{equation}
\sup_{i\geq0}\left\{  \sum\limits_{k=0}^{i}\left\vert \dfrac{\text{d}%
M(k)}{\text{d}x_{I,i}}\right\vert \right\}  \leq1+2C_{2}+2d_{2}\left(
C_{1}+C_{2}\right)  . \label{Inequ-2}%
\end{equation}

\end{The}

\textbf{Proof:} \ We first prove (\ref{Inequ-1}). To that end, it follows from
(\ref{In-1}) and (\ref{In-3}) that%
\begin{align*}
\sum\limits_{k=0}^{i}\left\vert \dfrac{\text{d}L(k)}{\text{d}x_{I,i}%
}\right\vert =  &  \sum\limits_{k=0}^{i-1}\left\vert \dfrac{\text{d}%
L(k)}{\text{d}x_{I,i}}\right\vert +\left\vert \dfrac{\text{d}L(k)}%
{\text{d}x_{I,i}}\right\vert \\
=  &  \sum\limits_{k=0}^{i-1}\left[  \left(  x_{I,i-1}+x_{W,k}\right)
^{d_{1}-1}-\left(  x_{I,i}+x_{W,k}\right)  ^{d_{1}-1}\right] \\
&  +\left(  d_{1}-1\right)  \sum\limits_{k=0}^{i}\left(  x_{I,i}%
-x_{I,i+1}\right)  \left(  x_{I,i}+x_{W,k}\right)  ^{d_{1}-2}\\
&  +\left(  x_{I,i}+x_{W,i}\right)  ^{d_{1}-1}.
\end{align*}
Note that%
\begin{align*}
&  \sum\limits_{k=0}^{i-1}\left[  \left(  x_{I,i-1}+x_{W,k}\right)  ^{d_{1}%
-1}-\left(  x_{I,i}+x_{W,k}\right)  ^{d_{1}-1}\right] \\
&  =\sum\limits_{k=0}^{i-1}\left(  x_{I,i-1}-x_{I,i}\right)  \sum_{n=0}%
^{d_{1}-2}\left(  x_{I,i-1}+x_{W,k}\right)  ^{n}\left(  x_{I,i}+x_{W,k}%
\right)  ^{d_{1}-2-n}\\
&  \leq\left(  d_{1}-1\right)  i\left(  x_{I,i-1}-x_{I,i}\right) \\
&  \leq\left(  d_{1}-1\right)  \sum_{i=1}^{\infty}i\left(  x_{I,i-1}%
-x_{I,i}\right) \\
&  =\left(  d_{1}-1\right)  \left(  x_{I,0}+\sum_{i=1}^{\infty}x_{I,i}\right)
\\
&  \leq\left(  d_{1}-1\right)  \left(  1+C_{1}\right)  ,
\end{align*}%
\begin{align*}
&  \left(  d_{1}-1\right)  \sum\limits_{k=0}^{i}\left(  x_{I,i}-x_{I,i+1}%
\right)  \left(  x_{I,i}+x_{W,k}\right)  ^{d_{1}-2}\\
&  \leq\left(  d_{1}-1\right)  \left(  x_{I,i}-x_{I,i+1}\right)
\sum\limits_{k=0}^{i}\left(  x_{I,i}+x_{W,k}\right)  ^{d_{1}-2}\\
&  \leq\left(  d_{1}-1\right)  \left(  i+1\right)  \left(  x_{I,i-1}%
-x_{I,i}\right) \\
&  \leq\left(  d_{1}-1\right)  \left(  \sum_{i=0}^{\infty}x_{I,i}%
+x_{I,0}\right) \\
&  \leq\left(  d_{1}-1\right)  \left(  2+C_{1}\right)
\end{align*}
and%
\[
\left(  x_{I,i}+x_{W,i}\right)  ^{d_{1}-1}\leq1,
\]
this gives that for any $i\geq0$%
\[
\sum\limits_{k=0}^{i}\left\vert \dfrac{\text{d}L(k)}{\text{d}x_{I,i}%
}\right\vert \leq1+\left(  d_{1}-1\right)  \left(  3+2C_{1}\right)  .
\]
Hence we have%
\[
\sup_{i\geq0}\left\{  \sum\limits_{k=0}^{i}\left\vert \dfrac{\text{d}%
L(k)}{\text{d}x_{I,i}}\right\vert \right\}  \leq1+\left(  d_{1}-1\right)
\left(  3+2C_{1}\right)  .
\]

In what follows we prove (\ref{Inequ-2}). Note that we will show that the key
condition: $d_{2}\geq2$, is due to the influence of the retrial rate $j\theta$
on our following computation.

It follows from (\ref{In-2}) and (\ref{In-4}) that%
\begin{align*}
\sum\limits_{k=0}^{i}\left\vert \dfrac{\text{d}M(k)}{\text{d}x_{I,i}%
}\right\vert =  &  \sum\limits_{k=0}^{i-1}\left\vert \dfrac{\text{d}%
M(k)}{\text{d}x_{I,i}}\right\vert +\left\vert \dfrac{\text{d}M(k)}%
{\text{d}x_{I,i}}\right\vert \\
=  &  \sum\limits_{k=0}^{i-1}\left\vert i\left(  x_{I,i}+x_{W,k}\right)
^{d_{2}-1}-\left(  i-1\right)  \left(  x_{I,i-1}+x_{W,k}\right)  ^{d_{2}%
-1}\right\vert \\
&  +\sum\limits_{k=0}^{i}\left(  d_{2}-1\right)  i\left(  x_{I,i}%
-x_{I,i+1}\right)  \left(  x_{I,i}+x_{W,k}\right)  ^{d_{2}-2}\\
&  +i\left(  x_{I,i}+x_{W,i}\right)  ^{d_{2}-1}.
\end{align*}
If $i\left(  x_{I,i}+x_{W,k}\right)  ^{d_{2}-1}\geq\left(  i-1\right)  \left(
x_{I,i-1}+x_{W,k}\right)  ^{d_{2}-1}$, then%
\[
\left\vert i\left(  x_{I,i}+x_{W,k}\right)  ^{d_{2}-1}-\left(  i-1\right)
\left(  x_{I,i-1}+x_{W,k}\right)  ^{d_{2}-1}\right\vert \leq x_{I,i}+x_{W,k};
\]
If $i\left(  x_{I,i}+x_{W,k}\right)  ^{d_{2}-1}<\left(  i-1\right)  \left(
x_{I,i-1}+x_{W,k}\right)  ^{d_{2}-1}$, then%
\[
\left\vert i\left(  x_{I,i}+x_{W,k}\right)  ^{d_{2}-1}-\left(  i-1\right)
\left(  x_{I,i-1}+x_{W,k}\right)  ^{d_{2}-1}\right\vert \leq i\left(
x_{I,i-1}-x_{I,i}\right)  .
\]
Thus we obtain%
\begin{align*}
&  \sum\limits_{k=0}^{i-1}\left\vert i\left(  x_{I,i}+x_{W,k}\right)
^{d_{2}-1}-\left(  i-1\right)  \left(  x_{I,i-1}+x_{W,k}\right)  ^{d_{2}%
-1}\right\vert \\
&  \leq\max\left\{  \sum\limits_{k=0}^{i-1}\left(  x_{I,i}+x_{W,k}\right)
,\sum\limits_{k=0}^{i-1}i\left(  x_{I,i-1}-x_{I,i}\right)  \right\} \\
&  =\max\left\{  ix_{I,i}+\sum\limits_{k=0}^{i-1}x_{W,k},i^{2}\left(
x_{I,i-1}-x_{I,i}\right)  \right\} \\
&  \leq\max\left\{  \sum\limits_{k=1}^{\infty}kx_{I,k}+\sum\limits_{k=0}%
^{\infty}x_{W,k},\sum\limits_{i=1}^{\infty}i^{2}\left(  x_{I,i-1}%
-x_{I,i}\right)  \right\} \\
&  =\max\left\{  \sum\limits_{k=1}^{\infty}kx_{I,k}+\sum\limits_{k=1}^{\infty
}x_{W,k}+x_{W,0},x_{I,0}+\sum\limits_{k=1}^{\infty}x_{I,k}+2\sum
\limits_{k=1}^{\infty}kx_{I,k}\right\} \\
&  \leq\max\left\{  1+C_{1}+C_{2},1+C_{1}+2C_{2}\right\} \\
&  =1+C_{1}+2C_{2},
\end{align*}%
\begin{align*}
&  \sum\limits_{k=0}^{i}\left(  d_{2}-1\right)  i\left(  x_{I,i}%
-x_{I,i+1}\right)  \left(  x_{I,i}+x_{W,k}\right)  ^{d_{2}-2}\\
&  \leq\left(  d_{2}-1\right)  i\left(  i+1\right)  \left(  x_{I,i}%
-x_{I,i+1}\right) \\
&  \leq\left(  d_{2}-1\right)  \sum\limits_{i=1}^{\infty}i\left(  i+1\right)
\left(  x_{I,i}-x_{I,i+1}\right) \\
&  \leq2\left(  d_{2}-1\right)  \sum\limits_{i=1}^{\infty}\left(  i+1\right)
x_{I,i}\\
&  \leq2\left(  d_{2}-1\right)  \left(  C_{1}+C_{2}\right)  ,
\end{align*}%
\[
i\left(  x_{I,i}+x_{W,i}\right)  ^{d_{2}-1}\leq i\left(  x_{I,i}%
+x_{W,i}\right)  \leq\sum_{i=1}^{\infty}i\left(  x_{I,i}+x_{W,i}\right)
\leq2C_{2},
\]
this gives that for any $i\geq0$%
\[
\sum\limits_{k=0}^{i}\left\vert \dfrac{\text{d}M(k)}{\text{d}x_{I,i}%
}\right\vert \leq1+2C_{2}+2d_{2}\left(  C_{1}+C_{2}\right)  .
\]
Thus we obtain%
\[
\sup_{i\geq0}\left\{  \sum\limits_{k=0}^{i}\left\vert \dfrac{\text{d}%
M(k)}{\text{d}x_{I,i}}\right\vert \right\}  \leq1+2C_{2}+2d_{2}\left(
C_{1}+C_{2}\right)  .
\]
This completes the proof. \textbf{{\rule{0.08in}{0.08in}}}

For any real matrix $\mathbf{D}=\left(  d_{i,j}\right)  _{0\leq i,j<\infty}$,
we define its norm as%
\[
\left\Vert \mathbf{D}\right\Vert =\sup_{i\geq0}\left\{  \sum_{j=0}^{\infty
}\left\vert d_{i,j}\right\vert \right\}  .
\]
At the same time, we introduce the notation%
\[
\left\vert D\right\vert =\left(  \left\vert d_{i,j}\right\vert \right)
_{0\leq i,j<\infty}.
\]
Hence it follows from (\ref{Equ5.10}) that%
\begin{equation}
\left\Vert \mathcal{D}F(x)\right\Vert =\sup_{i\geq0}\left\{  \left\Vert
\sum\limits_{j=0}^{i+1}\left\vert A_{i,j}\right\vert \right\Vert \right\}  .
\label{equ6.1}%
\end{equation}
If $\rho<1$ and $d_{2}\geq2$, then it follows from (\ref{A-1}) to (\ref{A-3})
that%
\begin{align*}
\left\Vert \mathcal{D}F(x)\right\Vert \leq &  \max\left\{  \lambda\sup
_{i\geq0}\left\{  \sum\limits_{k=0}^{i}\left\vert \dfrac{\text{d}%
L(k)}{\text{d}x_{I,i}}\right\vert \right\}  +\theta\sup_{i\geq0}\left\{
\sum\limits_{k=0}^{i}\left\vert \dfrac{\text{d}M(k)}{\text{d}x_{I,i}%
}\right\vert \right\}  ,\right. \\
&  \left.  \mu+\lambda\sup_{i\geq0}\left\{  \left\vert \dfrac{\text{d}%
L(i)}{\text{d}x_{I,i}}\right\vert \right\}  +\theta\sup_{i\geq0}\left\{
\left\vert \dfrac{\text{d}M(k)}{\text{d}x_{I,i}}\right\vert \right\}  ,\right.
\\
&  \left.  \mu+2\lambda d_{1}+\lambda\sup_{i\geq0}\left\{  \left\vert
\dfrac{\text{d}L(i)}{\text{d}x_{I,i}}\right\vert \right\}  +\theta\sup
_{i\geq0}\left\{  \left\vert \dfrac{\text{d}M(k)}{\text{d}x_{I,i}}\right\vert
\right\}  \right\} \\
\leq &  \mu+2\lambda d_{1}+\lambda\sup_{i\geq0}\left\{  \sum\limits_{k=0}%
^{i}\left\vert \dfrac{\text{d}L(k)}{\text{d}x_{I,i}}\right\vert \right\}
+\theta\sup_{i\geq0}\left\{  \sum\limits_{k=0}^{i}\left\vert \dfrac
{\text{d}M(k)}{\text{d}x_{I,i}}\right\vert \right\}  ,
\end{align*}
using Theorem \ref{The:Upper}, we obtain%
\[
\left\Vert \mathcal{D}F\left(  x\right)  \right\Vert \leq M,
\]
where%
\begin{align*}
M=  &  \mu+2\lambda d_{1}+\lambda\left[  1+\left(  d_{1}-1\right)  \left(
3+2C_{1}\right)  \right] \\
&  +\theta\left[  1+2C_{2}+2d_{2}\left(  C_{1}+C_{2}\right)  \right]  .
\end{align*}

Note that $x=\mathbf{u},$ this gives that for $\mathbf{u}\in\widetilde{\Omega
}$%
\begin{equation}
\left\Vert \mathcal{D}F\left(  \mathbf{u}\right)  \right\Vert \leq M.
\label{equ6-4}%
\end{equation}
It follows from (\ref{Eq7.3.3}) and (\ref{equ6-4}) that
\begin{equation}
\left\Vert F(\mathbf{u})-F(\mathbf{v})\right\Vert \leq M\left\Vert
(\mathbf{u}-\mathbf{v})\right\Vert \text{.} \label{equ7-1}%
\end{equation}
This indicates that the function $F(\mathbf{u})$ is Lipschitz for
$\mathbf{u}\in\widetilde{\Omega}$.

Note that $x=\mathbf{u}$, it follows from Equations (\ref{Equ5.8}) and
(\ref{Equ5.9}) that for $\mathbf{u}\in\widetilde{\Omega}$%
\[
\mathbf{u}\left(  t\right)  =\mathbf{u}\left(  0\right)  +\int_{0}^{t}F\left(
\mathbf{u}\left(  \xi\right)  \right)  \text{d}\xi,
\]
this gives%
\begin{equation}
\mathbf{u}\left(  t\right)  =\left(  \mathbf{g},\mathbf{h}\right)  +\int
_{0}^{t}F\left(  \mathbf{u}\left(  \xi\right)  \right)  \text{d}\xi.
\label{EquK-15}%
\end{equation}

Using the Picard approximation as well as the Lipschitz condition given in
(\ref{equ7-1}), it is easy to prove that there exists the unique solution to
the integral equation (\ref{EquK-15}) according to the basic results of the
Banach space. Therefore, there exists the unique solution to the system of
differential equations (\ref{Equ5.8}) and (\ref{Equ5.9}) (that is, the system
of differential equations (\ref{Equat13-0}) to (\ref{Equat17})) for $t\geq0$.

\begin{Rem}
If $d_{2}=1$, then for $k\geq0$%
\[
M\left(  k\right)  =\sum_{j=k}^{\infty}j\left(  x_{I,j}-x_{I,j+1}\right)  .
\]
This gives%
\[
\dfrac{\text{d}M(k)}{\text{d}x_{I,i}}=\left\{
\begin{array}
[c]{cc}%
1, & 0\leq k\leq i-1,\\
i, & k=i,\\
0 & k\geq i+1.
\end{array}
\right.
\]
We obtain%
\[
\sum\limits_{k=0}^{i}\left\vert \dfrac{\text{d}M(k)}{\text{d}x_{I,i}%
}\right\vert =2i,
\]
which leads to%
\[
\sup_{i\geq0}\left\{  \sum\limits_{k=0}^{i}\left\vert \dfrac{\text{d}%
M(k)}{\text{d}x_{I,i}}\right\vert \right\}  =+\infty.
\]

\end{Rem}

The following theorem provides an important property for the solution
$\mathbf{u}(t)=\mathbf{u}\left(  t,\mathbf{g},\mathbf{h}\right)  $, and
illustrates how this solution depends on the initial condition $\mathbf{u}%
(0)=\left(  \mathbf{g},\mathbf{h}\right)  .$

\begin{The}
Let $\mathbf{u}(t)$ be the unique and global solution to the system of
equations (\ref{Equat13-0}) to (\ref{Equat17}) for $t\geq0$, where
$\mathbf{u}(t)=\left(  u_{0}\left(  t\right)  ,u_{1}\left(  t\right)
,u_{2}\left(  t\right)  ,\ldots\right)  $ and $u_{k}\left(  t\right)  =\left(
u_{I,k}\left(  t\right)  ,u_{W,k}\left(  t\right)  \right)  $ for $k\geq0$.
Then%
\begin{equation}
u_{W,k}\left(  t\right)  \leq\sum_{i=1}^{k}u_{W,i}\left(  0\right)
\frac{\left(  \lambda t\right)  ^{k-i}}{\left(  k-i\right)  !}+\frac{2\left(
\lambda+\theta\right)  }{\lambda}\frac{\left(  \lambda t\right)  ^{k}}%
{k!}+\frac{\lambda+2\theta}{\lambda}\sum_{i=1}^{k}\frac{\left(  \lambda
t\right)  ^{i}}{i!} \label{equ8-1}%
\end{equation}
and%
\begin{align}
u_{I,k}\left(  t\right)  \leq &  u_{I,k}\left(  0\right)  +\frac{\mu}{\lambda
}\sum_{i=1}^{k}u_{W,i}\left(  0\right)  \frac{\left(  \lambda t\right)
^{k+1-i}}{\left(  k+1-i\right)  !}\nonumber\\
&  +\frac{2\mu\left(  \lambda+\theta\right)  }{\lambda^{2}}\frac{\left(
\lambda t\right)  ^{k+1}}{\left(  k+1\right)  !}+\frac{\mu\left(
\lambda+2\theta\right)  }{\lambda^{2}}\sum_{i=1}^{k}\frac{\left(  \lambda
t\right)  ^{i+1}}{\left(  i+1\right)  !}. \label{equ8-2}%
\end{align}
Furthermore, if $\left(  \mathbf{g},\mathbf{h}\right)  \in\Omega$, then
$\mathbf{u}\left(  t,\mathbf{g},\mathbf{h}\right)  \in\Omega$ for $t\geq0$.
\end{The}

\textbf{Proof: }We first prove Inequalities (\ref{equ8-1}) by induction. Then
we prove (\ref{equ8-2}) by means of Inequalities (\ref{equ8-1}).

For $l=1$, we have%
\begin{align*}
\frac{\text{d}u_{W,1}\left(  t\right)  }{\text{d}t}  &  \leq\lambda\left(
u_{W,0}\left(  t\right)  \right)  ^{d_{1}}+\lambda u_{I,1}\left(
u_{I,1}+u_{W,1}\right)  ^{d_{1}-1}+2\theta u_{I,2}\left(  u_{I,2}%
+u_{W,1}\right)  ^{d_{2}-1}\\
&  \leq\lambda u_{W,0}\left(  t\right)  +\left(  \lambda+2\theta\right)  ,
\end{align*}
this gives%
\begin{align*}
u_{W,1}\left(  t\right)   &  =u_{W,1}\left(  0\right)  +\int_{0}^{t}%
\text{d}u_{W,1}\left(  s\right) \\
&  \leq u_{W,1}\left(  0\right)  +\int_{0}^{t}\left[  \lambda u_{W,0}\left(
s\right)  +\left(  \lambda+2\theta\right)  \right]  \text{d}s\\
&  \leq u_{W,1}\left(  0\right)  +2\left(  \lambda+\theta\right)  t.
\end{align*}

For $l=2$, we have%
\[
\frac{\text{d}u_{W,2}\left(  t\right)  }{\text{d}t}\leq\lambda u_{W,1}\left(
t\right)  +\left(  \lambda+2\theta\right)  ,
\]
\begin{align*}
u_{W,2}\left(  t\right)   &  =u_{W,2}\left(  0\right)  +\int_{0}^{t}%
\text{d}u_{W,2}\left(  s\right)  \leq u_{W,2}\left(  0\right)  +\int_{0}%
^{t}\left[  \lambda u_{W,1}\left(  t\right)  +\left(  \lambda+2\theta\right)
\right]  \text{d}s\\
&  \leq\sum_{i=1}^{2}u_{W,i}\left(  0\right)  \frac{\left(  \lambda t\right)
^{2-i}}{\left(  2-i\right)  !}+\frac{2\left(  \lambda+\theta\right)  }%
{\lambda}\frac{\left(  \lambda t\right)  ^{2}}{2!}+\frac{\lambda+2\theta
}{\lambda}\sum_{i=1}^{2}\frac{\left(  \lambda t\right)  ^{i}}{i!}.
\end{align*}

We assume that for $l=k$, Inequality (\ref{equ8-1}) holds. In this case, we
shall prove that for $l=k+1$, Inequality (\ref{equ8-1}) also holds.

Since $u_{W,k}\left(  t\right)  \leq u_{W,k-1}\left(  t\right)  \leq1,$ it
follows from (\ref{Equat13-0}) that%
\[
\frac{\text{d}u_{W,k+1}\left(  t\right)  }{\text{d}t}\leq\lambda\left[
u_{W,k}\left(  t\right)  \right]  ^{d_{1}}+\left(  \lambda+2\theta\right)
\leq\lambda u_{W,k}\left(  t\right)  +\left(  \lambda+2\theta\right)  ,
\]
this gives%
\begin{align*}
u_{W,k+1}\left(  t\right)   &  =u_{W,k+1}\left(  0\right)  +\int_{0}%
^{t}\text{d}u_{W,k+1}\left(  s\right)  \leq u_{W,k+1}\left(  0\right)
+\int_{0}^{t}\left[  \lambda u_{W,k}\left(  t\right)  +\left(  \lambda
+2\theta\right)  \right]  \text{d}s\\
&  \leq\sum_{i=1}^{k+1}u_{W,i}\left(  0\right)  \frac{\left(  \lambda
t\right)  ^{k+1-i}}{\left(  k+1-i\right)  !}+\frac{2\left(  \lambda
+\theta\right)  }{\lambda}\frac{\left(  \lambda t\right)  ^{k+1}}{\left(
k+1\right)  !}+\frac{\lambda+2\theta}{\lambda}\sum_{i=1}^{k+1}\frac{\left(
\lambda t\right)  ^{i}}{i!}.
\end{align*}

On the other hand, it follows from (\ref{Equat14}) that%
\[
\frac{\text{d}u_{I,k}\left(  t\right)  }{\text{d}t}\leq\mu u_{W,k}\left(
t\right)  ,
\]
which follows%
\begin{align*}
u_{I,k}\left(  t\right)  \leq &  u_{I,k}\left(  0\right)  +\int_{0}%
^{t}\text{d}u_{I,k}\left(  s\right)  \leq u_{I,k}\left(  0\right)  +\int
_{0}^{t}\mu u_{W,k}\left(  s\right)  \text{d}s\\
\leq &  u_{I,k}\left(  0\right)  +\frac{\mu}{\lambda}\sum_{i=1}^{k}%
u_{W,i}\left(  0\right)  \frac{\left(  \lambda t\right)  ^{k+1-i}}{\left(
k+1-i\right)  !}\\
&  +\frac{2\mu\left(  \lambda+\theta\right)  }{\lambda^{2}}\frac{\left(
\lambda t\right)  ^{k+1}}{\left(  k+1\right)  !}+\frac{\mu\left(
\lambda+2\theta\right)  }{\lambda^{2}}\sum_{i=1}^{k}\frac{\left(  \lambda
t\right)  ^{i+1}}{\left(  i+1\right)  !}.
\end{align*}

Finally, it is easy to see from (\ref{equ8-1}) and (\ref{equ8-2}) that if
$\left(  \mathbf{g},\mathbf{h}\right)  \in\Omega$, then $\mathbf{u}\left(
t,\mathbf{g},\mathbf{h}\right)  \in\Omega$ for $t\geq0$. This completes the
proof. \textbf{{\rule{0.08in}{0.08in}}}

\section{The Chaos of Propagation}

In this section, we use the operator semigroup to provide a mean-field limit
(or chaos of propagation), and show that $\left\{  U\left(  t\right)
,t\geq0\right\}  $ is the limiting process of the sequence $\left\{
U^{\left(  N\right)  }\left(  t\right)  ,t\geq0\right\}  $ of Markov processes
who asymptotically approaches a single trajectory identified by the unique and
global solution $\mathbf{u}\left(  t\right)  $ to the infinite-dimensional
system of limiting differential equations (\ref{Equat13-0}) to (\ref{Equat17}).

For the vector $\mathbf{u}^{\left(  N\right)  }=\left(  u_{0}^{\left(
N\right)  },u_{1}^{\left(  N\right)  },u_{2}^{\left(  N\right)  }%
,\ldots\right)  $ where the size of the row vector $u_{k}^{\left(  N\right)
}$ is $2$ for $k\geq0$, we write%
\begin{align*}
\widetilde{\Omega}_{N}=  &  \left\{  \mathbf{u}^{\left(  N\right)  }:\left(
1,1\right)  \geq u_{0}^{\left(  N\right)  }\geq u_{1}^{\left(  N\right)  }\geq
u_{2}^{\left(  N\right)  }\geq\cdots\geq0,\right. \\
&  \left.  \text{ }Nu_{k}^{\left(  N\right)  }\text{ \ is a vector of
nonnegative integers for }k\geq0\right\}  .
\end{align*}
and%
\[
\Omega_{N}=\left\{  \mathbf{u}^{\left(  N\right)  }\in\widetilde{\Omega}%
_{N}:\mathbf{u}^{\left(  N\right)  }e<+\infty\right\}  .
\]
At the same time, for the vector $\mathbf{u}=\left(  u_{0},u_{1},u_{2}%
,\ldots\right)  $ where the size of the row vector $u_{k}$ is $2$ for $k\geq
0$, we set%
\[
\widetilde{\Omega}=\{\mathbf{u}:\left(  1,1\right)  \geq u_{0}\geq u_{1}\geq
u_{2}\geq\cdots\geq0\}
\]
and%
\[
\Omega=\left\{  \mathbf{u}\in\widetilde{\Omega}:\mathbf{u}e<+\infty\right\}
.
\]
Obviously, $\Omega_{N}\subsetneqq\Omega\subsetneqq\widetilde{\Omega}$ and
$\Omega_{N}\subsetneqq\widetilde{\Omega}_{N}\subsetneqq\widetilde{\Omega}$.

In the vector space $\widetilde{\Omega}$, we take a metric%
\begin{equation}
\rho\left(  \mathbf{u},\mathbf{u}^{\prime}\right)  =\max_{j=1,2}\left\{
\sup_{k\geq0}\left\{  \dfrac{|u_{k,j}-u_{k,j}^{\prime}|}{k+1}\right\}
\right\}  , \label{Equat8}%
\end{equation}
for $\mathbf{u},\mathbf{u}^{\prime}\in\widetilde{\Omega}$. Note that under the
metric $\rho\left(  \mathbf{u},\mathbf{u}^{\prime}\right)  ,$ the vector space
$\widetilde{\Omega}$ is separable and compact.

Now, we consider the sequence $\left\{  U^{\left(  N\right)  }\left(
t\right)  ,t\geq0\right\}  $ of Markov processes on state space $\widetilde
{\Omega}_{N}$ (or $\Omega_{N}$) for $N=1,2,3,\ldots$. Note that the stochastic
evolution of this retrial supermarket model of $N$ identical servers is
described as the Markov process $\left\{  U^{\left(  N\right)  }\left(
t\right)  ,t\geq0\right\}  $, where%
\[
\frac{\text{d}}{\text{d}t}U^{\left(  N\right)  }\left(  t\right)
=\mathbf{A}_{N}\text{ }f(U^{\left(  N\right)  }\left(  t\right)  ),
\]
$\mathbf{A}_{N}$ acting on functions $f:\Omega_{N}\rightarrow\mathbf{C}^{1}$
is the generating operator of the Markov process $\left\{  U^{\left(
N\right)  }\left(  t\right)  ,t\geq0\right\}  $, and%
\begin{equation}
\mathbf{A}_{N}=\mathbf{A}_{N}^{\text{Primary-In}}+\mathbf{A}_{N}%
^{\text{Retrial-In}}+\mathbf{A}_{N}^{\text{Out}}, \label{Equat9}%
\end{equation}
for $\mathbf{u=}\left(  \mathbf{g,h}\right)  \in\Omega_{N}$ with
$\mathbf{g}=\left(  g_{0},g_{1},g_{2},\ldots\right)  $ and $\mathbf{h}=\left(
h_{0},h_{1},h_{2},\ldots\right)  ,$%
\begin{align*}
\mathbf{A}_{N}^{\text{Primary-In}}=  &  \lambda N\left\{  \sum_{k=1}^{\infty
}\left(  g_{k-1}^{d_{1}}-g_{k}^{d_{1}}\right)  \left[  f\left(  \mathbf{g}%
+\frac{\mathbf{e}_{k\text{ }}}{N},\mathbf{h}\right)  -f\left(  \mathbf{g,h}%
\right)  \right]  \right. \\
&  +\left.  \sum_{k=0}^{\infty}\sum_{j=k}^{\infty}\left(  h_{j}-h_{j+1}%
\right)  \left(  h_{j}+g_{k}\right)  ^{d_{1}-1}\left[  f\left(  \mathbf{g}%
+\frac{\mathbf{e}_{j\text{ }}}{N},\mathbf{h}\right)  -f\left(  \mathbf{g,h}%
\right)  \right]  \right\}  ,
\end{align*}
\[
\mathbf{A}_{N}^{\text{Retrial-In}}=N\theta\sum_{k=0}^{\infty}\sum
_{j=k+1}^{\infty}j\left(  h_{j}-h_{j+1}\right)  \left(  h_{j}+g_{k}\right)
^{d_{2}-1}\left[  f\left(  \mathbf{g}+\frac{\mathbf{e}_{j\text{ }}}%
{N},\mathbf{h}-\frac{\mathbf{e}_{j\text{ }}}{N}\right)  -f\left(
\mathbf{g,h}\right)  \right]  ,
\]%
\begin{equation}
\mathbf{A}_{N}^{\text{Out}}=\mu N\sum_{k=1}^{\infty}g_{k}\left[  f\left(
\mathbf{g}-\frac{\mathbf{e}_{k\text{ }}}{N},\mathbf{h}\right)  -f\left(
\mathbf{g,h}\right)  \right]  ,\nonumber
\end{equation}
where $\mathbf{e}_{k\text{ }}$stands for a row vector with the $k$th entry be
one and all the other entries be zero. Therefore, for $(\mathbf{g}%
,\mathbf{h})\in\Omega_{N}$ and the function $f:\Omega_{N}\rightarrow
\mathbf{C}^{1}$, we obtain%
\begin{align}
\mathbf{A}_{N}f\left(  \mathbf{g,h}\right)  =  &  \lambda N\left\{  \sum
_{k=1}^{\infty}\left(  g_{k-1}^{d_{1}}-g_{k}^{d_{1}}\right)  \left[  f\left(
\mathbf{g}+\frac{\mathbf{e}_{k\text{ }}}{N},\mathbf{h}\right)  -f\left(
\mathbf{g,h}\right)  \right]  \right. \nonumber\\
&  +\left.  \sum_{k=0}^{\infty}\sum_{j=k}^{\infty}\left(  h_{j}-h_{j+1}%
\right)  \left(  h_{j}+g_{k}\right)  ^{d_{1}-1}\left[  f\left(  \mathbf{g}%
+\frac{\mathbf{e}_{j\text{ }}}{N},\mathbf{h}\right)  -f\left(  \mathbf{g,h}%
\right)  \right]  \right\} \nonumber\\
&  +N\theta\sum_{k=0}^{\infty}\sum_{j=k+1}^{\infty}j\left(  h_{j}%
-h_{j+1}\right)  \left(  h_{j}+g_{k}\right)  ^{d_{2}-1}\left[  f\left(
\mathbf{g}+\frac{\mathbf{e}_{j\text{ }}}{N},\mathbf{h}-\frac{\mathbf{e}%
_{j\text{ }}}{N}\right)  -f\left(  \mathbf{g,h}\right)  \right] \nonumber\\
&  -\mu N\sum_{k=1}^{\infty}g_{k}\left[  f\left(  \mathbf{g},\mathbf{h}%
-\frac{\mathbf{e}_{k\text{ }}}{N}\right)  -f\left(  \mathbf{g,h}\right)
\right]  . \label{Equat10}%
\end{align}

The operator semigroup of the Markov process $\left\{  U^{\left(  N\right)
}\left(  t\right)  ,t\geq0\right\}  $ is defined as $\mathbf{T}_{N}(t)$. If
$f:\Omega_{N}\rightarrow\mathbf{C}^{1}$, then for $\left(  \mathbf{g,h}%
\right)  \in\Omega_{N}$ and $t\geq0$%
\begin{equation}
\mathbf{T}_{N}(t)f(\mathbf{g,h})=E\left[  f(U^{\left(  N\right)  }\left(
t\right)  )\text{ }|\text{ }U^{\left(  N\right)  }\left(  0\right)  =\left(
\mathbf{g,h}\right)  \right]  . \label{Equat11}%
\end{equation}
Note that $\mathbf{A}_{N}$ is the generating operator of the operator
semigroup $\mathbf{T}_{N}(t)$, it is easy to see that $\mathbf{T}_{N}%
(t)=\exp\left\{  \mathbf{A}_{N}t\right\}  $ for $t\geq0$.

Let $L=C(\widetilde{\Omega})$ be the Banach space of continuous functions
$f:\widetilde{\Omega}\rightarrow\mathbf{C}^{1}$ with uniform metric
$\left\Vert f\right\Vert =\underset{\mathbf{u}\in\widetilde{\Omega}}{\max
}\left\vert f(\mathbf{u})\right\vert $, and similarly, let $L_{N}=C(\Omega
_{N})$. The inclusion $\Omega_{N}\subset\widetilde{\Omega}$ induces a
contraction mapping $\Pi_{N}:L\rightarrow L_{N},\Pi_{N}f(\mathbf{u}%
)=f(\mathbf{u})$ for $f\in L$ and $\mathbf{u}\in\Omega_{N}$.

Now, we consider the limiting behavior of the sequence $\{U^{\left(  N\right)
}\left(  t\right)  $, $t\geq0\}$ of the Markov processes for $N=1,2,3,\ldots$.
Two formal limits for the sequence $\left\{  \mathbf{A}_{N}\right\}  $ of the
generating operators and for the sequence $\left\{  \mathbf{T}_{N}(t)\right\}
$ of the semigroups are expressed as $\mathbf{A}=\lim_{N\rightarrow\infty
}\mathbf{A}_{N}$ and $\mathbf{T}\left(  t\right)  =\lim_{N\rightarrow\infty
}\mathbf{T}_{N}(t)$ for $t\geq0$, respectively. It follows from (\ref{Equat10}%
) that as $N\rightarrow\infty$%
\begin{align}
\mathbf{A}f(\mathbf{g},\mathbf{h})=  &  \lambda\sum_{k=1}^{\infty}\left(
g_{k-1}^{d_{1}}-g_{k}^{d_{1}}\right)  \frac{\partial}{\partial g_{k}%
}f(\mathbf{g},\mathbf{h})\nonumber\\
&  +\lambda\sum_{k=0}^{\infty}\sum_{j=k}^{\infty}\left(  h_{j}-h_{j+1}\right)
\left(  h_{j}+g_{k}\right)  ^{d_{1}-1}\frac{\partial}{\partial g_{j}%
}f(\mathbf{g},\mathbf{h})\nonumber\\
&  +\theta\sum_{k=0}^{\infty}\sum_{j=k+1}^{\infty}j\left(  h_{j}%
-h_{j+1}\right)  \left(  h_{j}+g_{k}\right)  ^{d_{2}-1}\left[  \frac{\partial
}{\partial g_{j}}f(\mathbf{g},\mathbf{h})-\frac{\partial}{\partial h_{j}%
}f(\mathbf{g},\mathbf{h})\right] \nonumber\\
&  -\mu\sum_{k=1}^{\infty}g_{k}\frac{\partial}{\partial g_{k}}f(\mathbf{g}%
,\mathbf{h}). \label{Equat12}%
\end{align}

We define a mapping: $\left(  \mathbf{g,h}\right)  \rightarrow\mathbf{u}%
(t,\mathbf{g,h})$, where $\mathbf{u}(t,\mathbf{g,h})$ is a solution to the
system of differential equations (\ref{Equat13-0}) to (\ref{Equat17}) for
$t\geq0$. Note that the operator semigroup $\mathbf{T}(t)$ acts in the space
$L$. If $f\in L$ and $(\mathbf{g,h})\in\widetilde{\Omega}$, then%
\begin{equation}
\mathbf{T}(t)f(\mathbf{g,h})=f(\mathbf{u}(t,\mathbf{g,h})). \label{Equat18}%
\end{equation}
It is easy to see that the operator semigroups $\mathbf{T}_{N}(t)$ and
$\mathbf{T}(t)$ are strongly continuous and constructive, see, for example,
Section 1.1 in Chapter one of Ethier and Kurtz \cite{Eth:1986}. We denote by
$\mathcal{D}(\mathbf{A})$ the domain of the generating operator $\mathbf{A}$.
It follows from (\ref{Equat18}) that if $f$ is a function from $L$ and has the
partial derivatives $\dfrac{\partial}{\partial g_{k}}f(\mathbf{g,h}%
),\dfrac{\partial}{\partial h_{k}}f(\mathbf{g,h})\in L$ for $k\geq0$, and
$\underset{k\geq0}{\sup}\left\{  \left\vert \dfrac{\partial}{\partial g_{k}%
}f(\mathbf{g,h})\right\vert ,\left\vert \dfrac{\partial}{\partial h_{k}%
}f(\mathbf{g,h})\right\vert \right\}  <\infty$, then $f\in\mathcal{D}%
(\mathbf{A})$.

Let $D$ be the set of all functions $f\in L$ that have the partial derivatives
$\dfrac{\partial}{\partial g_{k}}f(\mathbf{g,h})$, $\dfrac{\partial}{\partial
h_{k}}f(\mathbf{g,h}),\dfrac{\partial^{2}}{\partial g_{i}\partial g_{j}%
}f(\mathbf{g,h}),\dfrac{\partial^{2}}{\partial g_{i}\partial h_{j}%
}f(\mathbf{g,h})$ and $\dfrac{\partial^{2}}{\partial h_{i}\partial h_{j}%
}f(\mathbf{g,h})$, and there exists $C=C(f)<+\infty$ such that%
\begin{equation}
\sup_{\substack{k\geq0\\(\mathbf{g,h})\in\widetilde{\Omega}}}\left\{  \left|
\dfrac{\partial}{\partial g_{k}}f(\mathbf{g,h})\right|  ,\left|
\dfrac{\partial}{\partial h_{k}}f(\mathbf{g,h})\right|  \right\}  <C
\label{Eq10.7}%
\end{equation}
and%
\begin{equation}
\sup_{\substack{i,j\geq0\\(\mathbf{g,h})\in\widetilde{\Omega}}}\left\{
\left|  \dfrac{\partial^{2}}{\partial g_{i}\partial g_{j}}f(\mathbf{g,h}%
)\right|  ,\left|  \dfrac{\partial^{2}}{\partial g_{i}\partial h_{j}%
}f(\mathbf{g,h})\right|  ,\left|  \dfrac{\partial^{2}}{\partial h_{i}\partial
h_{j}}f(\mathbf{g,h})\right|  \right\}  <C. \label{Eq10.8}%
\end{equation}

We call that $f\in L$ depends only on the first $K+1$ two-dimensional
variables if for $(\mathbf{g}^{\left(  1\right)  }\mathbf{,h}^{\left(
1\right)  }),(\mathbf{g}^{\left(  2\right)  }\mathbf{,h}^{\left(  2\right)
})\in\widetilde{\Omega},$ it follows from $g_{i}^{\left(  1\right)  }%
=g_{i}^{\left(  2\right)  }$ and $h_{i}^{\left(  1\right)  }=h_{i}^{\left(
2\right)  }$ for $0\leq i\leq K$ that $f(\mathbf{g}^{\left(  1\right)
}\mathbf{,h}^{\left(  1\right)  })=f(\mathbf{g}^{\left(  2\right)
}\mathbf{,h}^{\left(  2\right)  })$. A similar and simple proof to that in
Proposition 2 in Vvedenskaya et al \cite{Vve:1996} can show that the set of
functions from $L$ that depends on the first finite two-dimensional variables
is dense in $L$.

The following lemma comes from Proposition 1 in Vvedenskaya et al
\cite{Vve:1996}. We restate it here for the convenience of description.

\begin{Pro}
\label{Pro:inf}Consider an infinite-dimensional linear system of differential
equations%
\[
\frac{\text{d}z_{k}\left(  t\right)  }{\text{d}t}=\sum_{i=0}^{\infty}%
z_{i}\left(  t\right)  a_{i,k}\left(  t\right)  +b_{k}\left(  t\right)
,\text{ \ }k=0,1,2,\ldots,t\geq0,
\]
let $\sum_{i=0}^{\infty}\left|  a_{k,i}\left(  t\right)  \right|  \leq a,$
$b_{k}\left(  t\right)  \leq b_{0}\exp\left\{  bt\right\}  \ $and
$z_{k}\left(  t\right)  \leq\varrho,$ where $b_{0}\geq0$ and $a<b$. Then for
$k=0,1,2,\ldots,$%
\[
z_{k}\left(  t\right)  \leq\varrho\exp\left\{  at\right\}  +\frac{b_{0}}%
{b-a}\left[  \exp\left\{  bt\right\}  -\exp\left\{  at\right\}  \right]  .
\]

\end{Pro}

Let%
\[
M_{1}=\theta\sum_{j=1}^{\infty}u_{I,j},
\]%
\[
M_{2}=\theta\sum_{j=1}^{\infty}ju_{I,j}%
\]
and%
\[
M_{3}=\theta\sum_{j=1}^{\infty}ju_{W,j}.
\]
Then $M_{1}\leq\theta C_{1}$ and $M_{2},M_{3}\leq\theta C_{2}$.

\begin{Lem}
\label{Lem:Bound}If the vector $\mathbf{u}(t)=\mathbf{u}(t,\mathbf{g,h})$ is a
solution to the system of differential equations (\ref{Equat13-0}) to
(\ref{Equat17}) for $t\geq0$, then%
\begin{align}
&  \max\left\{  \left\vert \frac{\partial u_{W,k}\left(  t,\mathbf{g,h}%
\right)  }{\partial g_{j}}\right\vert ,\left\vert \frac{\partial
u_{W,k}\left(  t,\mathbf{g,h}\right)  }{\partial h_{j}}\right\vert \right\}
\nonumber\\
&  \leq\exp\left\{  \left(  3\lambda d_{1}+\mu+2\lambda+\left(  d_{2}%
+1\right)  M_{1}+M_{2}+M_{3}\right)  t\right\}  , \label{equ10-1}%
\end{align}%
\begin{align}
&  \max\left\{  \left\vert \frac{\partial u_{I,k}\left(  t,\mathbf{g,h}%
\right)  }{\partial g_{j}}\right\vert ,\left\vert \frac{\partial
u_{I,k}\left(  t,\mathbf{g,h}\right)  }{\partial h_{j}}\right\vert \right\}
\nonumber\\
&  \leq\exp\left\{  \left(  \mu+\lambda\left(  3d_{1}-2\right)  +\theta
ld_{2}+\left(  2d_{2}-1\right)  M_{1}+M_{2}+M_{3}\right)  t\right\}  ,
\label{equ10-2}%
\end{align}%
\begin{align}
&  \max\left\{  \left\vert \dfrac{\partial^{2}u_{W,k}\left(  t,\mathbf{g,h}%
\right)  }{\partial g_{i}g_{j}}\right\vert ,\left\vert \dfrac{\partial
^{2}u_{W,k}\left(  t,\mathbf{g,h}\right)  }{\partial g_{i}h_{j}}\right\vert
,\left\vert \dfrac{\partial^{2}u_{W,k}\left(  t,\mathbf{g,h}\right)
}{\partial h_{i}h_{j}}\right\vert \right\} \nonumber\\
&  \leq\frac{3\lambda d_{1}+\mu+2\lambda+\left(  d_{2}+1\right)  M_{1}%
+M_{2}+M_{3}}{\lambda\left(  4d_{1}^{2}-6d_{1}+3\right)  +\left(
d_{2}-1\right)  \left(  d_{2}-2\right)  \left(  M_{1}+M_{2}\right)  }\left[
\exp\left\{  2ax\right\}  -\exp\left\{  ax\right\}  \right]  , \label{equ10-3}%
\end{align}
and%
\begin{align}
&  \max\left\{  \left\vert \dfrac{\partial^{2}u_{I,k}\left(  t,\mathbf{g,h}%
\right)  }{\partial g_{i}g_{j}}\right\vert ,\left\vert \dfrac{\partial
^{2}u_{I,k}\left(  t,\mathbf{g,h}\right)  }{\partial g_{i}h_{j}}\right\vert
,\left\vert \dfrac{\partial^{2}u_{I,k}\left(  t,\mathbf{g,h}\right)
}{\partial h_{i}h_{j}}\right\vert \right\} \nonumber\\
&  \leq\frac{\mu}{\mu+\lambda\left(  3d_{1}-2\right)  +\theta ld_{2}+\left(
2d_{2}-1\right)  M_{1}+M_{2}+M_{3}}\exp\left\{  2bx\right\}  -\exp\left\{
bx\right\}  . \label{equ10-4}%
\end{align}

\end{Lem}

\textbf{Proof: }We only prove (\ref{equ10-1}), while the other three
inequalities (\ref{equ10-2})\ to (\ref{equ10-4}) can be proved similarly.

It is easy to verify that the vector $\mathbf{u}(t)=\mathbf{u}(t,\mathbf{g,h}%
)$ possesses the following derivatives%
\[
\frac{\partial u_{W,k}\left(  t,\mathbf{g,h}\right)  }{\partial g_{j}}%
,\frac{\partial u_{W,k}\left(  t,\mathbf{g,h}\right)  }{\partial h_{j}}%
,\frac{\partial u_{I,k}\left(  t,\mathbf{g,h}\right)  }{\partial g_{j}}%
,\frac{\partial u_{I,k}\left(  t,\mathbf{g,h}\right)  }{\partial h_{j}},
\]%
\begin{align*}
&  \dfrac{\partial^{2}u_{W,k}\left(  t,\mathbf{g,h}\right)  }{\partial
g_{i}\partial g_{j}},\dfrac{\partial^{2}u_{W,k}\left(  t,\mathbf{g,h}\right)
}{\partial g_{i}\partial h_{j}},\dfrac{\partial^{2}u_{W,k}\left(
t,\mathbf{g,h}\right)  }{\partial h_{i}\partial h_{j}},\\
&  \dfrac{\partial^{2}u_{I,k}\left(  t,\mathbf{g,h}\right)  }{\partial
g_{i}\partial g_{j}},\dfrac{\partial^{2}u_{I,k}\left(  t,\mathbf{g,h}\right)
}{\partial g_{i}\partial h_{j}},\dfrac{\partial^{2}u_{I,k}\left(
t,\mathbf{g,h}\right)  }{\partial h_{i}\partial h_{j}}.
\end{align*}
For all $k,i\geq0$, we write $u_{W,k;i}^{\prime}=\frac{\partial u_{W,k}\left(
t,\mathbf{g,h}\right)  }{\partial g_{i}}$ and $u_{I,k;i}^{\prime}%
=\frac{\partial u_{I,k}\left(  t,\mathbf{g,h}\right)  }{\partial g_{i}}$. Then
it follows from (\ref{Equat13-0}) that the sequence $\left\{  u_{W,k;j}%
^{\prime}\right\}  $ satisfies the following differential equation%
\begin{align*}
\frac{\text{d}u_{W,k;i}^{\prime}}{\text{d}t}=  &  \lambda d_{1}\left(
u_{W,k-1}\right)  ^{d_{1}-1}u_{W,k-1;i}^{\prime}+\left[  -\lambda d_{1}\left(
u_{W,k}\right)  ^{d_{1}-1}-\mu+\lambda\sum_{j=k}^{\infty}\left(
u_{I,j}-u_{I,j+1}\right)  \left(  d_{1}-1\right)  \right. \\
&  \left.  \times\left(  u_{I,j}+u_{W,k}\right)  ^{d_{1}-2}+\theta\sum
_{j=k+1}^{\infty}j\left(  u_{I,j}-u_{I,j+1}\right)  \left(  d_{2}-1\right)
\left(  u_{I,j}+u_{W,k}\right)  ^{d_{2}-2}u_{W,k;i}^{\prime}\right. \\
&  \left.  +\lambda\left(  u_{I,k}+u_{W,k}\right)  ^{d_{1}-1}+\lambda
\sum_{j=k}^{\infty}\left(  u_{I,j}-u_{I,j+1}\right)  \left(  d_{1}-1\right)
\left(  u_{I,j}+u_{W,k}\right)  ^{d_{1}-2}\right]  u_{I,k;i}^{\prime}\\
&  +\sum_{j=k+1}^{\infty}\left\{  [\lambda\left(  u_{I,j}+u_{W,k}\right)
^{d_{1}-1}-\lambda\left(  u_{I,j-1}+u_{W,k}\right)  ^{d_{1}-1}+\lambda\left(
u_{I,j}-u_{I,j+1}\right)  \left(  d_{1}-1\right)  \right. \\
&  \left.  \left.  \times\left(  u_{I,k+2}+u_{W,k}\right)  ^{d_{1}-2}+\theta
j\left(  u_{I,j}+u_{W,k}\right)  ^{d_{2}-1}d_{2}\left(  u_{I,j}-u_{I,j+1}%
\right)  +u_{I,j+1}+u_{W,k}\right]  u_{I,j;i}^{\prime}\right\}  .
\end{align*}
Applying proposition \ref{Pro:inf} to the solution of the above differential
equation with $a=3\lambda d_{1}+\mu+2\lambda+\left(  d_{2}+1\right)
M_{1}+M_{2}+M_{3},$ $b_{0}=0,$ $\varrho=1$, we obtain the first inequality
(\ref{equ10-1}). This completes the proof. \textbf{{\rule{0.08in}{0.08in}}}

\begin{Lem}
\label{Lem:Core}The set$\mathcal{\ }D$ is a core for the generating operator
$\mathbf{A}$.
\end{Lem}

\textbf{Proof: }It is obvious that $D$ is dense in $L$ and $D\in
\mathcal{D}(\mathbf{A})$. Let $D_{0}$ be the set of functions from $D$ which
depend only on the first finite two-dimensional variables. It is easy to see
that $D_{0}$ is dense in $L$. Using proposition 3.3 in Chapter 1 of Ethier and
Kurtz \cite{Eth:1986}, it can show that for any $t\geq0$ the operator
semigroup $\mathbf{T}(t)$ does not bring $D_{0}$ out of $D$. Select an
arbitrary function $\varphi\in D_{0}$ and let $f(\mathbf{g},\mathbf{h}%
)=\varphi(\mathbf{u}(t,\mathbf{g},\mathbf{h}))$ for $(\mathbf{g}%
,\mathbf{h)}\in\widetilde{\Omega}$. It follows from Lemma \ref{Lem:Bound} that
$f$ has the partial derivatives%
\[
\frac{\partial u_{W,k}\left(  t,\mathbf{g,h}\right)  }{\partial g_{j}}%
,\frac{\partial u_{W,k}\left(  t,\mathbf{g,h}\right)  }{\partial h_{j}}%
,\frac{\partial u_{I,k}\left(  t,\mathbf{g,h}\right)  }{\partial g_{j}}%
,\frac{\partial u_{I,k}\left(  t,\mathbf{g,h}\right)  }{\partial h_{j}},
\]%
\[
\dfrac{\partial^{2}u_{W,k}\left(  t,\mathbf{g,h}\right)  }{\partial
g_{i}\partial g_{j}},\dfrac{\partial^{2}u_{W,k}\left(  t,\mathbf{g,h}\right)
}{\partial g_{i}\partial h_{j}},\dfrac{\partial^{2}u_{W,k}\left(
t,\mathbf{g,h}\right)  }{\partial h_{i}\partial h_{j}},
\]
and%
\[
\dfrac{\partial^{2}u_{I,k}\left(  t,\mathbf{g,h}\right)  }{\partial
g_{i}\partial g_{j}},\dfrac{\partial^{2}u_{I,k}\left(  t,\mathbf{g,h}\right)
}{\partial g_{i}\partial h_{j}},\dfrac{\partial^{2}u_{I,k}\left(
t,\mathbf{g,h}\right)  }{\partial h_{i}\partial h_{j}}%
\]
and they satisfy the inequalities (\ref{equ10-1}) to (\ref{equ10-3}).
Therefore $f\in D$. This completes the proof. \textbf{{\rule{0.08in}{0.08in}}}

The following theorem applies the operator semigroup to providing the
mean-field limiting process $\left\{  U\left(  t\right)  ,t\geq0\right\}  $
for the sequence $\left\{  U^{\left(  N\right)  }\left(  t\right)
,t\geq0\right\}  $ of Markov processes, and indicates that this sequence of
Markov processes asymptotically approaches a single trajectory identified by
the unique and global solution to the system of differential equations
(\ref{Equat13-0}) to (\ref{Equat17}) for $t\geq0$. Note that this proof is
based on Inequalities (\ref{Ineq-15}) and (\ref{Ineq-16}).

\begin{The}
\label{The:Lim}Let $f$ be continuous functions $f:\widetilde{\Omega
}\rightarrow\mathbf{C}^{1}$. Then for any $t>0$%
\[
\lim_{N\rightarrow\infty}\underset{(\mathbf{g},\mathbf{h)}\in\Omega_{N}}{\sup
}\left\vert \mathbf{T}_{N}\left(  t\right)  f\left(  \mathbf{g},\mathbf{h}%
\right)  -f\left(  \mathbf{u}(t,\mathbf{g},\mathbf{h})\right)  \right\vert
=0.
\]
The convergence is uniform in $t\in\left[  0,T\right]  $ for any $T>0.$
\end{The}

\textbf{Proof: }This proof is to use the convergence of the sequence of
operator semigroups as well as the convergence of the sequence of their
generating generators, e.g., see Theorem 6.1 in Chapter 1 of Ethier and Kurtz
\cite{Eth:1986}. Note that Lemma \ref{Lem:Core} shows that the
set$\mathcal{\ }D$ is a core for the generating operator $\mathbf{A}$, thus
for any function $f\in D$ we obtain%
\[
N\left[  f(\mathbf{g+}\frac{\mathbf{e}_{k}}{N},\mathbf{h})-f(\mathbf{g}%
,\mathbf{h})\right]  +\frac{\partial}{\partial g_{k}}f(\mathbf{g}%
,\mathbf{h})=-\frac{\gamma_{1,k}\left(  g\right)  }{N}\frac{\partial
^{2}f\left(  \mathbf{g}+\gamma_{2,k}\left(  g\right)  \frac{\mathbf{e}_{k}}%
{N},\mathbf{h}\right)  }{\partial g_{k}^{2}},
\]%
\[
N\left[  f(\mathbf{g},\mathbf{h+}\frac{\mathbf{e}_{k}}{N})-f(\mathbf{g}%
,\mathbf{h})\right]  +\frac{\partial}{\partial h_{k}}f(\mathbf{g}%
,\mathbf{h})=-\frac{\gamma_{1,k}\left(  h\right)  }{N}\frac{\partial
^{2}f\left(  \mathbf{g},\mathbf{h}+\gamma_{2,k}\left(  h\right)
\frac{\mathbf{e}_{k}}{N}\right)  }{\partial h_{k}^{2}}.
\]
where $0<\gamma_{i,k}\left(  g\right)  ,\gamma_{i,k}\left(  h\right)  <1$ for
$i=1,2$. Since there exists a constant $\eta>0$ such that%
\[
\left\vert \frac{\gamma_{1,k}\left(  g\right)  }{N}\frac{\partial^{2}f\left(
\mathbf{g}+\gamma_{2,k}\left(  g\right)  \frac{\mathbf{e}_{k}}{N}%
,\mathbf{h}\right)  }{\partial g_{k}^{2}}\right\vert \leq\frac{\eta}{N}%
\]
and%
\[
\left\vert \frac{\gamma_{1,k}\left(  h\right)  }{N}\frac{\partial^{2}f\left(
\mathbf{g},\mathbf{h}+\gamma_{2,k}\left(  h\right)  \frac{\mathbf{e}_{k}}%
{N}\right)  }{\partial h_{k}^{2}}\right\vert \leq\frac{\eta}{N},
\]
we obtain%
\begin{align*}
\left\vert \mathbf{A}_{N}f(\mathbf{g},\mathbf{h})-f(\mathbf{g},\mathbf{h}%
)\right\vert \leq &  \frac{\eta}{N}\left[  \lambda\sum_{k=1}^{\infty}\left(
g_{k-1}^{d_{1}}-g_{k}^{d_{1}}\right)  +\lambda\sum_{l=0}^{\infty}\sum
_{j=l}^{\infty}\left(  h_{j}-h_{j+1}\right)  \left(  h_{j}+g_{l}\right)
^{d_{1}-1}\right. \\
&  \left.  +\theta\sum_{k=0}^{\infty}\sum_{j=k+1}^{\infty}j\left(
h_{j}-h_{j+1}\right)  \left(  h_{j}+g_{k}\right)  ^{d_{2}-1}+\mu\sum
_{k=1}^{\infty}g_{k}\right]  .
\end{align*}
Note that $0<g_{i},h_{j}\leq1$, $g_{i}+h_{j}\leq g_{0}+h_{0}=1$, and it is
seen from Inequalities (\ref{Ineq-15}) and (\ref{Ineq-16}) that $\sum
_{j=1}^{\infty}g_{j}\leq C_{1}$, $\sum_{j=1}^{\infty}h_{j}\leq C_{1}$ and
$\sum_{j=1}^{\infty}jh_{j}\leq C_{2}$, thus we obtain%
\[
\sum_{k=1}^{\infty}\left(  g_{k-1}^{d_{1}}-g_{k}^{d_{1}}\right)  =g_{0}%
^{d_{1}}\leq1,
\]%
\begin{align*}
&  \sum_{l=0}^{\infty}\sum_{j=l}^{\infty}\left(  h_{j}-h_{j+1}\right)  \left(
h_{j}+g_{l}\right)  ^{d_{1}-1} =\sum_{j=0}^{\infty}\sum_{l=0}^{j}\left(
h_{j}-h_{j+1}\right)  \left(  h_{j}+g_{l}\right)  ^{d_{1}-1}\\
&  =\sum_{j=0}^{\infty}\left(  h_{j}-h_{j+1}\right)  \sum_{l=0}^{j}\left(
h_{j}+g_{l}\right)  ^{d_{1}-1} \leq\sum_{j=0}^{\infty}\left(  j+1\right)
\left(  h_{j}-h_{j+1}\right) \\
&  =h_{0}+\sum_{j=1}^{\infty}h_{j} \leq1+C_{1},
\end{align*}%
\begin{align*}
&  \sum_{k=0}^{\infty}\sum_{j=k}^{\infty}j\left(  h_{j}-h_{j+1}\right)
\left(  h_{j}+g_{k}\right)  ^{d_{2}-1} =\sum_{j=0}^{\infty}\sum_{k=0}%
^{j}j\left(  h_{j}-h_{j+1}\right)  \left(  h_{j}+g_{k}\right)  ^{d_{2}-1}\\
&  \leq\sum_{j=0}^{\infty}j\left(  j+1\right)  \left(  h_{j}-h_{j+1}\right)
=2\sum_{j=1}^{\infty}\left(  j+1\right)  h_{j} \leq2\left(  C_{1}%
+C_{2}\right)
\end{align*}
and%
\[
\sum_{j=1}^{\infty}g_{j}\leq C_{1},
\]
we obtain%
\begin{align*}
\left\vert \mathbf{A}_{N}f(\mathbf{g},\mathbf{h})-f(\mathbf{g},\mathbf{h}%
)\right\vert  &  \leq\frac{\eta}{N}\left[  \lambda\left(  g_{0}^{d_{1}}%
+\sum_{j=0}^{\infty}h_{j}\right)  +2\theta\sum_{j=1}^{\infty}\left(
j+1\right)  h_{j}+\mu\sum_{k=1}^{\infty}g_{k}\right] \\
&  \leq\frac{\eta}{N}\left[  \lambda\left(  2+C_{1}\right)  +2\theta\left(
C_{1}+C_{2}\right)  +\mu C_{1}\right]
\end{align*}
Note that $\eta$, $C_{1}$ and $C_{2}$ are all finite, it is clear that as
$N\rightarrow\infty$,%
\[
\lim_{N\rightarrow\infty}\underset{(\mathbf{g},\mathbf{h)}\in\Omega_{N}}{\sup
}\left\vert \mathbf{T}_{N}\left(  t\right)  f\left(  \mathbf{g},\mathbf{h}%
\right)  -f\left(  \mathbf{u}(t,\mathbf{g},\mathbf{h})\right)  \right\vert
=0.
\]
This completes this proof. \textbf{{\rule{0.08in}{0.08in}}}

If $\lim_{N\rightarrow\infty}U^{\left(  N\right)  }\left(  0\right)
=\mathbf{u}(0)=\left(  \mathbf{g},\mathbf{h}\right)  \in$ $\Omega$ in
probability, then $U\left(  t\right)  =\lim_{N\rightarrow\infty}U^{\left(
N\right)  }\left(  t\right)  $ is concentrated on the trajectory
$\Gamma_{\left(  \mathbf{g},\mathbf{h}\right)  }=\left\{  \mathbf{u}%
(t,\mathbf{g},\mathbf{h}):t\geq0\right\}  $. This indicates the functional
strong law of large numbers for the time evolution of the fractions of the
busy servers and of the idle servers, thus the sequence $\left\{  U^{\left(
N\right)  }\left(  t\right)  ,t\geq0\right\}  $ of Markov processes converges
weakly to the expected fraction vector $\mathbf{u}(t,\mathbf{g},\mathbf{h})$
as $N\rightarrow\infty$, that is, for any $T>0$%
\[
\lim_{N\rightarrow\infty}\sup_{0\leq s\leq T}\left\Vert U^{\left(  N\right)
}\left(  s\right)  -\mathbf{u}(s,\mathbf{g},\mathbf{h})\right\Vert =0\text{
\ in probability}.
\]
For the limits of stochastic process sequences, readers may refer to Chen and
Yao \cite{Chen:2001} and Whitt \cite{Whi:2002} for more details.

\section{Computation of the Fixed Point}

In this section, we compute the fixed point by means of a system of nonlinear
equations. Then we use the fixed point to provide performance analysis of this
retrial supermarket model through some numerical computation.

A row vector $\pi=\left(  \pi_{W,0},\pi_{I,0},\pi_{W,1},\pi_{I,1},\pi
_{W,2},\pi_{I,2},\ldots\right)  $ is called a fixed point of the system of
differential equations (\ref{Equat13-0}) to (\ref{Equat17}) satisfied by the
limiting expected fraction vector $\mathbf{u}\left(  t\right)  $ if $\pi
=\lim_{t\rightarrow+\infty}\mathbf{u}\left(  t\right)  $, that is, $\pi
_{W,k}=\lim_{t\rightarrow+\infty}u_{W,k}\left(  t\right)  $ and $\pi
_{I,k}=\lim_{t\rightarrow+\infty}u_{I,k}\left(  t\right)  $ for $k\geq0$. It
is well-known that if $\pi$ is a fixed point of the vector $\mathbf{u}\left(
t\right)  $, then%
\[
\lim_{t\rightarrow+\infty}\left[  \frac{\text{d}}{\text{d}t}\mathbf{u}\left(
t\right)  \right]  =0.
\]
To determine the fixed point $\pi$, as $t\rightarrow+\infty$ taking limits on
both sides of Equations (\ref{Equat13-0}) to (\ref{Equat16}) we obtain a
system of nonlinear equations as follows: For $k\geq1$%
\begin{align}
&  \left.  \lambda\left(  \pi_{W,k-1}^{d_{1}}-\pi_{W,k}^{d_{1}}\right)
-\mu\pi_{W,k}+\lambda\sum_{j=k}^{\infty}\left(  \pi_{I,j}-\pi_{I,j+1}\right)
\left(  \pi_{I,j}+\pi_{W,k}\right)  ^{d_{1}-1}\right. \nonumber\\
&  \left.  +\theta\sum_{j=k+1}^{\infty}j\left(  \pi_{I,j}-\pi_{I,j+1}\right)
\left(  \pi_{I,j}+\pi_{W,k}\right)  ^{d_{2}-1}=0\right.  , \label{FP0}%
\end{align}
and for $l\geq0$%
\begin{align}
&  \left.  \mu\pi_{W,l}-\lambda\sum_{j=l}^{\infty}\left(  \pi_{I,j}%
-\pi_{I,j+1}\right)  \left(  \pi_{I,j}+\pi_{W,l}\right)  ^{d_{1}-1}\right.
\nonumber\\
&  \left.  -\theta\sum_{j=l}^{\infty}j\left(  \pi_{I,j}-\pi_{I,j+1}\right)
\left(  \pi_{I,j}+\pi_{W,l}\right)  ^{d_{2}-1}=0\right.  , \label{FP1}%
\end{align}
\begin{equation}
\pi_{I,0}+\pi_{W,0}=1 \label{FP2}%
\end{equation}

Using (\ref{FP0}) and (\ref{FP1}) for $k,l\geq1$, we obtain%
\begin{equation}
\lambda\left(  \pi_{W,k-1}^{d_{1}}-\pi_{W,k}^{d_{1}}\right)  -\theta k\left(
\pi_{I,k}-\pi_{I,k+1}\right)  \left(  \pi_{I,k}+\pi_{W,k}\right)  ^{d_{2}%
-1}=0. \label{FP3}%
\end{equation}
This gives%
\begin{equation}
\lambda\pi_{W,0}^{d_{1}}-\theta\sum_{k=1}^{\infty.}k\left(  \pi_{I,k}%
-\pi_{I,k+1}\right)  \left(  \pi_{I,k}+\pi_{W,k}\right)  ^{d_{2}-1}=0
\label{FP4}%
\end{equation}

\begin{Rem}
Analysis of retrial queues has been an interesting and more challenging topic
in the area of queues for many years, where the main difficulties stem from
the computation of the stationary distribution of the number of customers in
the orbit through solving the systems of linear equations corresponding to the
level-dependent Markov chains, e.g., see Artalejo and G\'{o}mez-Corral
\cite{Art:2008} and Li \cite{Li:2010} for applications of the $RG$%
-factorizations to many retrial queues. In this retrial supermarket model, we
organize the more general system of nonlinear equations (\ref{FP0}) and
(\ref{FP2}). Because the $RG$-factorizations and the generating functions
cannot be applied to deal with the system of nonlinear equations, we believe
it is a key to develop some effective methods for solving such systems of
nonlinear equations in the future study of retrial supermarket models.
\end{Rem}

The following theorem provides two different expressions for computing the
boundary value $\pi_{W,0}$ (note that $\pi_{I,0}=1-\pi_{W,0}$), while its
proof is easy by means of (\ref{FP4}), (\ref{FP1}) for $l=0$ and (\ref{FP2}).

\begin{The}
The boundary value $\pi_{W,0}$ is given by%
\begin{equation}
\pi_{W,0}^{d_{1}}=\frac{\theta}{\lambda}\sum_{k=1}^{\infty.}k\left(  \pi
_{I,k}-\pi_{I,k+1}\right)  \left(  \pi_{I,k}+\pi_{W,k}\right)  ^{d_{2}-1}
\label{FixP1}%
\end{equation}
and%
\begin{align}
\pi_{W,0} =  &  \rho\sum_{j=0}^{\infty}\left(  \pi_{I,j}-\pi_{I,j+1}\right)
\left(  \pi_{I,j}+\pi_{W,0}\right)  ^{d_{1}-1}\nonumber\\
&  +\frac{\theta}{\lambda}\sum_{j=1}^{\infty}j\left(  \pi_{I,j}-\pi
_{I,j+1}\right)  \left(  \pi_{I,j}+\pi_{W,0}\right)  ^{d_{2}-1}. \label{FixP2}%
\end{align}

\end{The}

To understand the importance of (\ref{FixP1}) and (\ref{FixP2}), we consider
some special cases as follows:

\textbf{Case one} $d_{2}=1$: In this case, it follows from (\ref{FixP1}) that%
\begin{equation}
\pi_{W,0}^{d_{1}}=\frac{\theta}{\lambda}\sum_{k=1}^{\infty.}k\left(  \pi
_{I,k}-\pi_{I,k+1}\right)  =\frac{\theta}{\lambda}\sum_{k=1}^{\infty.}%
\pi_{I,k}, \label{FixP4}%
\end{equation}
this gives%
\begin{equation}
\pi_{W,0}=\sqrt[d_{1}]{\frac{\theta}{\lambda}\sum_{k=1}^{\infty.}\pi_{I,k}}.
\label{FixP3}%
\end{equation}

\textbf{Case two} $d_{1}=d_{2}=1$: Note that this case corresponds to the
M/M/1 retrial queue. In the M/M/1 retrial queue, from (2.16) of Kulkarni and
Liang \cite{Kul:1997} we have that $\pi_{W,0}=\rho$ and $\pi_{I,0}=1-\rho$. It
follows from (\ref{FixP4}) that%
\begin{equation}
\sum_{j=1}^{\infty}\pi_{I,j}=\frac{\lambda}{\theta}\rho.\label{FixP5}%
\end{equation}
It follows from (\ref{FP0}) and (\ref{FP1}) that for $k\geq1$%
\begin{equation}
\lambda\left(  \pi_{W,k-1}-\pi_{W,k}\right)  -\mu\pi_{W,k}+\lambda\pi
_{I,k}+\theta\left[  \left(  k+1\right)  \pi_{I,k+1}+\sum_{j=k+2}^{\infty}%
\pi_{I,j}\right]  =0\label{FP7}%
\end{equation}
and%
\begin{equation}
\mu\pi_{W,k}-\lambda\pi_{I,k}-\theta\left[  k\pi_{I,k}+\sum_{j=k+1}^{\infty
}\pi_{I,j}\right]  =0.\label{FP8}%
\end{equation}
It follows from (\ref{FP8}) for $k=1$ and (\ref{FixP5}) that%
\begin{equation}
\mu\pi_{W,1}-\lambda\pi_{I,1}=\lambda\rho.\label{FixP6}%
\end{equation}
It follows from (\ref{FP3}) that%
\[
\lambda\left(  \pi_{W,k-1}-\pi_{W,k}\right)  +k\theta\left(  \pi_{I,k+1}%
-\pi_{I,k}\right)  =0,
\]
this gives%
\begin{equation}
\frac{\pi_{W,k-1}-\pi_{W,k}}{\pi_{I,k}-\pi_{I,k+1}}=\frac{\theta}{\lambda
}k,\text{ \ }k\geq1,\label{FP10}%
\end{equation}
It follows from (\ref{FP8}) that%
\[
\mu\pi_{W,k+1}-\lambda\pi_{I,k+1}-\theta\left[  \left(  k+1\right)
\pi_{I,k+1}+\sum_{j=k+2}^{\infty}\pi_{I,j}\right]  =0,
\]
which, together with (\ref{FP7}), follows%
\begin{equation}
\lambda\left(  \pi_{W,k-1}-\pi_{W,k}\right)  -\mu\left(  \pi_{W,k}-\pi
_{W,k+1}\right)  +\lambda\left(  \pi_{I,k}-\pi_{I,k+1}\right)  =0.\label{FP13}%
\end{equation}
Using (\ref{FP10}) and (\ref{FP13}), we obtain%
\begin{equation}
\frac{\pi_{W,k-1}-\pi_{W,k}}{\pi_{W,k}-\pi_{W,k+1}}=\frac{k\theta}{\rho\left(
\lambda+k\theta\right)  },\text{ \ }k\geq1,\label{FP11}%
\end{equation}
It follows from (\ref{FP11}) that
\[
\rho\lambda\pi_{W,0}+\theta\rho\sum_{j=0}^{\infty}\pi_{W,j}=\theta\sum
_{j=1}^{\infty}\pi_{W,j},
\]
this gives%
\begin{equation}
\sum_{j=1}^{\infty}\pi_{W,j}=\frac{\rho^{2}\left(  \lambda+\theta\right)
}{\theta\left(  1-\rho\right)  }.\label{FP9.1}%
\end{equation}
Hence the mean of the stationary number of customers in the orbit is given by%
\[
\sum_{j=1}^{\infty}\left(  \pi_{W,j}+\pi_{I,j}\right)  =\frac{\lambda\rho
}{1-\rho}\left(  \frac{1}{\mu}+\frac{1}{\theta}\right)  .
\]

\begin{Rem}
Based on the fixed point $\pi$, it is easy to have a useful relation as
follows:%
\[
\lim_{t\rightarrow+\infty}\lim_{N\rightarrow\infty}\mathbf{u}^{\left(
N\right)  }(t,\mathbf{g,h})=\lim_{N\rightarrow\infty}\lim_{t\rightarrow
+\infty}\mathbf{u}^{\left(  N\right)  }(t,\mathbf{g,h})=\pi.
\]
Therefore, we obtain%
\[
\lim_{\substack{N\rightarrow\infty\\t\rightarrow+\infty}}\mathbf{u}^{\left(
N\right)  }(t,\mathbf{g,h})=\pi.
\]

\end{Rem}

In the remainder of this section, we provide an effectively approximate
algorithm for solving the system of nonlinear equations (\ref{FP0}),
(\ref{FP1}) and (\ref{FP2}). In this algorithm, we truncate the size of each
orbit to the finite capacity of size $M$. In this case, we obtain the system
of nonlinear equations as follows: For $1\leq k\leq M-1$%
\begin{align}
&  \left.  \lambda\left(  \pi_{W,k-1}^{d_{1}}-\pi_{W,k}^{d_{1}}\right)
-\mu\pi_{W,k}+\lambda\sum_{j=k}^{M}\left(  \pi_{I,j}-\pi_{I,j+1}\right)
\left(  \pi_{I,j}+\pi_{W,k}\right)  ^{d_{1}-1}\right. \nonumber\\
&  \left.  +\theta\sum_{j=k+1}^{M}j\left(  \pi_{I,j}-\pi_{I,j+1}\right)
\left(  \pi_{I,j}+\pi_{W,k}\right)  ^{d_{2}-1}=0\right.  , \label{TFP-1}%
\end{align}
for $k=M$%
\begin{equation}
\lambda\left(  \pi_{W,M-1}^{d_{1}}-\pi_{W,M}^{d_{1}}\right)  -\mu\pi
_{W,M}+\lambda\pi_{I,M}\left(  \pi_{I,M}+\pi_{W,M}\right)  ^{d_{1}-1}
\label{TFP-2}%
\end{equation}
for $0\leq l\leq M-1$%
\begin{align}
&  \left.  \mu\pi_{W,l}-\lambda\sum_{j=l}^{M}\left(  \pi_{I,j}-\pi
_{I,j+1}\right)  \left(  \pi_{I,j}+\pi_{W,l}\right)  ^{d_{1}-1}\right.
\nonumber\\
&  \left.  -\theta\sum_{j=l}^{M}j\left(  \pi_{I,j}-\pi_{I,j+1}\right)  \left(
\pi_{I,j}+\pi_{W,l}\right)  ^{d_{2}-1}=0\right.  , \label{TFP-3}%
\end{align}
for $l=M$%
\begin{equation}
\mu\pi_{W,M}-\lambda\pi_{I,M}\left(  \pi_{I,M}+\pi_{W,M}\right)  ^{d_{1}%
-1}-\theta M\pi_{I,M}\left(  \pi_{I,M}+\pi_{W,M}\right)  ^{d_{2}-1}=0,
\label{TFP-4}%
\end{equation}%
\begin{equation}
\pi_{I,0}+\pi_{W,0}=1 \label{TFP-5}%
\end{equation}
and%
\begin{equation}
\pi_{I,M+1}=0. \label{TFP-6}%
\end{equation}

It is easy to check that the nonlinear equations (\ref{TFP-1}) to
(\ref{TFP-6}) for $d_{1}=d_{2}=1$ is the same as the corresponding tail
equations in the M/M/1/M retrial queue. Also, we show that it is convenient to
apply the MATLAB to solving the system of nonlinear equations (\ref{TFP-1}) to
(\ref{TFP-6}) numerically. Based on this, the mean of the number of stationary
customers in any orbit of this supermarket model is approximately given by%
\[
E\left[  Q\right]  =\sum_{k=1}^{\infty}\left(  \pi_{I,k}+\pi_{W,k}\right)
\approx\sum_{k=1}^{M}\left(  \pi_{I,k}+\pi_{W,k}\right)  ,
\]
when the truncated number $M$ is sufficiently large. In fact, when $d_{1}%
\geq2$ or $d_{2}\geq2$, the two sequences $\left\{  \pi_{I,k}\right\}  $ and
$\left\{  \pi_{W,k}\right\}  $ monotonically decrease to zero under a
super-exponential decay. Thus, when the truncated number is chosen to $M=50$,
our approximate computation for $E\left[  Q\right]  $ will arrive at a higher precision.

In what follows we consider two numerical examples, which are used to show
that our approximate algorithm is effective in the study of retrial
supermarket models.

\textbf{Example one: The role of arrival processes}

In the retrial supermarket models, we take that the exponential service rate
$\mu=7$ and the exponential retrial rate $\theta=2$. When $\rho<1$, it is
clear that the Poisson arrival rate $\lambda\in\left(  0,7\right)  $. Figure 4
shows that $E\left[  Q\right]  $ increases as $\lambda$ increases for any pair
$\left(  d_{1},d_{2}\right)  $ with $d_{1},d_{2}=1,2,5$. At the same time, it
is seen that $E\left[  Q\right]  $ decreases very fast as $d_{1}$ and $d_{2}$\ increase.

\begin{figure}[ptbh]
\centering
\includegraphics[width=11cm]{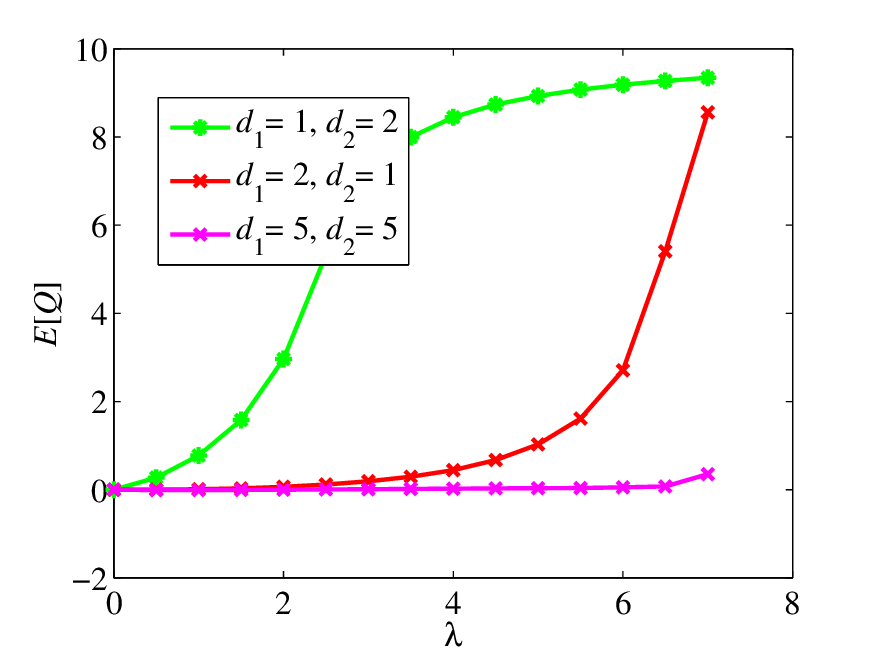} \caption{The mean $E\left[  Q\right]
$ vs $\lambda$ for some different pairs $\left(  d_{1},d_{2}\right)  $}%
\label{figure: fig-4}%
\end{figure}

\textbf{Example two: The role of retrial processes}

In the retrial supermarket models, we take that the Poisson arrive rate
$\lambda=2$ and the exponential service rate $\mu=3$. Clearly, $\rho<1$. Let
the retrial rate $\theta\in\left(  0,7\right)  $. Figure 5 shows that
$E\left[  Q\right]  $ decreases as $\theta$ increases for any pair $\left(
d_{1},d_{2}\right)  $ with $d_{1},d_{2}=1,2,5$. Also, it is clear that
$E\left[  Q\right]  $ decreases very fast as $d_{1}$ and $d_{2}$\ increase.

\begin{figure}[ptbh]
\centering
\includegraphics[width=11cm]{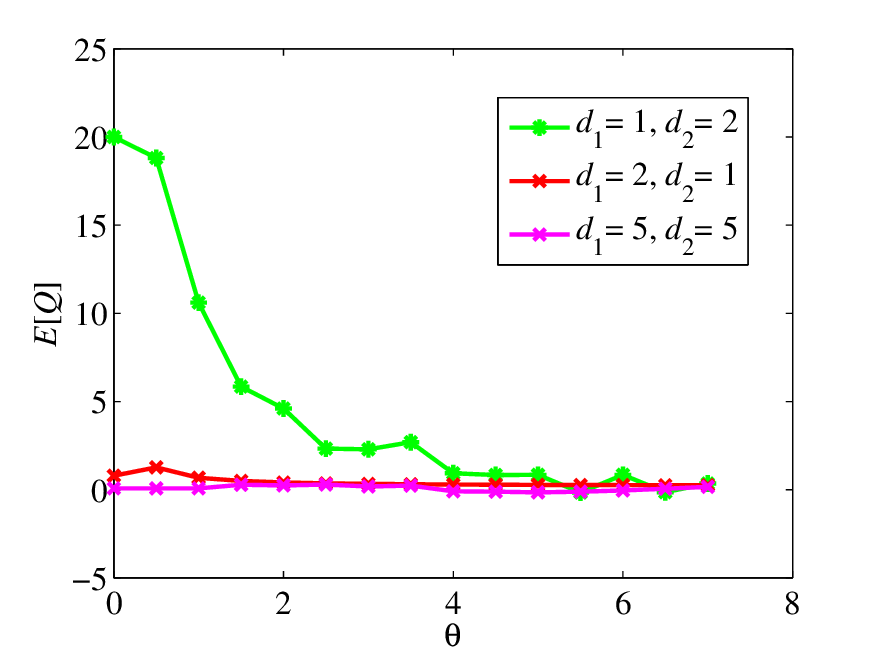} \caption{The mean $E\left[  Q\right]
$ vs $\theta$ for some different pairs $\left(  d_{1},d_{2}\right)  $}%
\label{figure: fig-5}%
\end{figure}

\section{Concluding remarks}

In this paper, we introduce and study a class of interesting retrial queueing
networks: the retrial supermarket models. Our main results provide a clear
picture for illustrating how to apply the mean-field theory to the study of
retrial supermarket models. This picture is organized by three basic steps as
follows. Step one: Providing a detailed probability computation to set up a
system of differential equations satisfied by the expected fraction vector
(see Section 3). Step two: Giving strictly mathematical proofs for the
mean-field limit as well as establishing the Lipschitz condition (see Sections
4 and 5). Step three: Computing the fixed point and analyzing performance
measures of this system (see Section 6).

Our mean-field method given in this paper is useful and effective in the study
of retrial supermarket models with applications to, such as data centers,
multi-core server systems and cloud computational modeling. We expect that
this mean-field method will be applicable to analyzing more general retrial
supermarket models, with characteristics such as, non-Poisson arrival
processes, non-exponential service times, and interesting random factors (for
example, server breakdowns and repairs, server vacations, negative customers
and impatient customers). Because there are few available works on the
analysis of retrial queueing networks in the current literature, we believe
the mean-field method given in this paper can open a new avenue in the future
study of retrial supermarket models, and more generally, of retrial queueing networks.

\section*{Acknowledgements}

The authors are grateful to Professor Henk C. Tijms whose comments and
suggestions greatly help us to improve the presentation of this paper. Q.L. Li
and Y. Wang thank that this research is partly supported by the National
Natural Science Foundation of China (No. 71271187, No. 61001075) and the Hebei
Natural Science Foundation of China (No. A2012203125).

Our research on the retrial supermarket models began from 2010. We have
obtained many valuable and constructive comments and suggestions from
Professor Jesus R. Artalejo. Therefore, we would like to report this work in
memory of our great friend: Jesus R. Artalejo.

\end{document}